\RequirePackage{ifpdf}
\input epsf.tex
\epsfclipon
\documentclass[11pt,a4paper,amsart,nofootinbib]{article}
\usepackage{jheppub}
\usepackage{epsfig,multicol}
\usepackage{multirow}
\usepackage{subcaption}
\usepackage[]{graphicx}
\usepackage[T1]{fontenc} 
\usepackage{cancel}
\usepackage[framemethod=TikZ]{mdframed}
\usepackage{lipsum}
\allowdisplaybreaks
\usepackage{stackengine}
\usepackage{tikz}
\usepackage{soul}

\usepackage[utf8]{inputenc}
 
\usepackage{ytableau}
 \ytableausetup{centertableaux,smalltableaux}
 
\usepackage[mathscr]{euscript}

\stackMath

\usepackage{hyperref}
\usepackage{multirow,tabularx}

\usepackage{cleveref}
\usepackage{mathtools}
\usepackage{longtable}
\usepackage{stackengine}
\usepackage{color}

\usepackage{epsfig}
\usepackage{epstopdf}
\usepackage{epsf}
\usepackage{ytableau}
 \ytableausetup{centertableaux}
\input{epsf.sty}
\usepackage{graphicx,amsmath,amssymb}
\usepackage{graphicx}
\usepackage{epsfig,multicol}
\usepackage{multirow}
\allowdisplaybreaks
\hypersetup{
  colorlinks=true,
  citecolor=magenta,
  linkcolor=blue,
  urlcolor=violet
 }
\usepackage{tikz} 

\usepackage{bbm,bm,amsmath,amssymb}
 
\usepackage[normalem]{ulem}

\usepackage{comment}
\def\be{\begin{equation}}
\def\ee{\end{equation}}
\def\figs/B{B}
\def\bea{\begin{eqnarray}}
\def\eea{\end{eqnarray}}
\def\bg{\begin{eqnarray}}
\def\nd{\end{eqnarray}}
\def\sin{{\rm sin}}
\def\cos{{\rm cos}}

\def\log{{\rm log}}
\def\ln{{\rm log}}

\def\be{\begin{equation}}
\def\ee{\end{equation}}

\def\doi{http://doi.org}

\def\Xint#1{\mathchoice
{\XXint\displaystyle\textstyle{#1}}%
{\XXint\textstyle\scriptstyle{#1}}%
{\XXint\scriptstyle\scriptscriptstyle{#1}}%
{\XXint\scriptscriptstyle\scriptscriptstyle{#1}}%
\!\int}
\def\XXint#1#2#3{{\setbox0=\hbox{$#1{#2#3}{\int}$ }
\vcenter{\hbox{$#2#3$ }}\kern-.6\wd0}}

\def\dashint{\Xint-}

\usepackage{physics}
\usepackage{tikz}
\usetikzlibrary{arrows.meta}
\usetikzlibrary{angles,quotes} 
\usetikzlibrary{decorations.markings,arrows.meta}
\tikzset{>=latex} 

\tikzset{
  midarr/.style={decoration={markings,mark=at position #1 with {\arrow{stealth}}},postaction={decorate}},
  midarr/.default=0.5
}
\colorlet{xcol}{blue!70!black}

\title{\boldmath $U(N)$ Torus Link Invariants in the Large $N$ limit from Matrix Model Approach}

\author[1]{Archana Maji,} \emailAdd{archana\_phy@iitb.ac.in}
\author[2,3]{Kushal Chakraborty,}\emailAdd{kushal16@iiserb.ac.in}
\author[2]{Suvankar Dutta,}\emailAdd{suvankar@iiserb.ac.in}
\author[1]{P. Ramadevi}\emailAdd{ramadevi@iitb.ac.in}

\affiliation[1]{Department of Physics, Indian Institute of Technology Bombay\\
Mumbai-400076, India }
\affiliation[2]{Indian Institute of Science Education and Research Bhopal \\
Bhopal bypass, Bhopal 462066, India}
\affiliation[3]{Mandelstam Institute for Theoretical Physics, School of Physics, University of the Witwatersrand, Johannesburg, WITS 2050, South Africa}

\abstract{ In this paper we study $U(N)$ colored HOMFLY-PT polynomials of torus links in the \textit{double scaling limit}  (polynomial variable $q\rightarrow 1$, $N\rightarrow \infty$ keeping $q^N$ fixed). We show that, in this limit, the colored HOMFLY-PT polynomial of any $(L\alpha,L\beta)$ torus link can be expressed in terms of the colored HOMFLY-PT polynomial of $(L,L)$ torus link. Using the connection between matrix models and the Chern-Simons field theoretic invariants, we show that the colored torus link invariants are uniquely expressed in terms of connected correlation functions of operators in $U(N)$ matrix model.  We  determine the leading and  subleading contribution to some of the correlators at large $N$ from the matrix model approach and find that they match exactly with those obtained from the corresponding colored HOMFLY-PT polynomials.

}
\begin{document} 
\maketitle
\flushbottom

\section{Introduction}
\label{sec:intro}

The famous discovery of Jones polynomial \cite{jones1985polynomial,jones1987hecke} by Vaughan Jones in 1984 led to a rejuvenation of the mathematical  theory of knots and links. A knot $\mathcal{K}$ is an embedding of a circle $S^1$ inside a three-manifold $M$ whereas a link $\mathcal{L}$ is a collection of two or more knots linked in a nontrivial way. Two knots $\mathcal{K}_{1}$ and $\mathcal{K}_{2}$ are considered to be equivalent if one can be deformed to another through any of the three Reidemeister moves. The central problem of knot theory is to classify different knots and links. To address this classification problem there comes various polynomial  invariants viz Alexander polynomial \cite{alexander1928topological}, Jones polynomial, HOMFLY-PT polynomial \cite{freyd1985new,przytycki2016invariants} \textit{et cetra} in the order of increasing sophistication.

Knot theory has become of spectacular interest to physicists ever since Witten's seminal work giving an intrinsically three dimensional definition of Jones polynomial from the perspective of quantum field theory \cite{witten1989quantum}. The Chern-Simons theory, a three dimensional gauge theory with an underlying gauge group $G$ provides a natural framework to compute the topological invariants of three-manifolds and of knots and links embedded inside those three-manifolds. In particular, the correlation function of the observables of Chern-Simons theory, viz.  Wilson loop operators, is nothing but Jones polynomial when the gauge group at consideration is $SU(2)$ and the representations put on the Wilson loops are that of the fundamental representation $\bigg(R\equiv \text{\ydiagram[]{1}}\bigg)$ of $SU(2)$. Likewise, with the fundamental representations of $G=SU(N)$ and $SO(N)$ we obtain HOMFLY-PT polynomial and Kauffman polynomial \cite{kauffman1990invariant} respectively. If other higher dimensional representations are placed on the Wilson loops then we obtain the corresponding colored polynomials. 

In this work we confine ourselves to a class of links called torus links that can be embedded on the surface of a torus in such a way that they do not cross over themselves. Any $L$ component torus link can be denoted as $(L\alpha,L\beta)$ where $L\alpha$ and $L\beta$ denote the number of times the link wraps around the meridian and the longitude of the torus respectively. Here $\alpha$ and $\beta$ are coprime to each other. These torus links can be obtained as a closure of $L\alpha$ strand braid with $(L\alpha -1)L\beta$ crossings.  We will study these torus link invariants in the double scaling limit of Chern-Simons theory associated with gauge group $U(N)$. This limit is defined as follows
\begin{align}\label{eq:dslimit1}
    N,k\rightarrow \infty~~~\text{such that}~~\lambda=\frac{N}{N+k}=\textrm{fixed} ,
\end{align}
where $k$ is the coupling constant of the Chern-Simons theory. In terms of the variable $q=\exp \left(\frac{2\pi i}{k+N} \right)$ the above limit can be rephrased as 
\begin{align}
   q\rightarrow 1, N\rightarrow \infty~~~\textrm{such that}~q^{N} =\textrm{fixed}.\label{eq:dslimit2}
\end{align}
By two of the authors, it has been shown that in this limit the invariants for a specific set of torus links viz. $(2,2m)$ torus links can be expressed in terms of $(2,2)$ torus link (also called Hopf link) invariant \cite{chakraborty2022large}. Interestingly, this result can be generalised for any arbitrary $L$ component torus link. In particular, we show that in the double scaling limit we can express the $(L\alpha,L\beta)$ torus link invariant  in terms of $(L,L)$ torus link invariant. 

Although the partition function of Chern-Simons theory can be obtained for some three-manifolds, the calculation of correlation functions is difficult in general. However, in the double scaling limit one can use the saddle point approximation to compute some of the correlation functions. In particular, the correlation function that corresponds to $(2,2)$ torus link invariant  was computed in Ref. \cite{chakraborty2022large} by mapping the problem to that of an incompressible fluid with intial and final fluid densities being related to the representations placed on the two component knots. In this paper we further study the correlation functions corresponding to $(L,L)$ torus link in the double scaling limit, for any arbitrary $L$.  We obtain an exact expression for the leading contribution of such link invariants using the saddle point approximation \cite{Chakraborty:2021oyq,chakraborty2022large}. We validate this result by explicitly computing the colored HOMFLY-PT polynomials for $(3,3)$ torus link carrying same/different representations on its three components. 

 It is  well known in knot theory, that all knot invariants reduce to unknot invariant in the double scaling limit. Similarly any $L $ component link invariant reduces to the product of $L$ unknot invariants. The intertwining  between the components is not captured in this limit. These observations are also seen in the  matrix model results using saddle point approximation.  Specifically, the leading order contribution to any $n$-point correlator in the large $N$ limit is the product of $n$ one-point correlators. Hence, we need to investigate the subleading corrections to capture the intertwining between the component knots constituting the link.
 The subleading corrections to the knot invariants are captured by the connected piece of correlation functions in the matrix model approach. We use the techniques developed in Ref. \cite{eynard2015random, Ambjorn:1992gw} to compute the connected pieces. We then validate the results for the subleading contributions of Hopf link invariants for some specific representations placed on the components of the link. Further, we show that \emph{any} knot or link invariant modulo the leading contribution (i.e. subtracting the leading contribution from the link invariant) can be uniquely written in terms of connected correlation function of operators in $U(N)$ matrix model. These connected correlators are related to the \textit{reformulated invariants} (which are written in terms of the  colored HOMFLY-PT polynomials)\cite{ramadevi2012reformulated}.

The organisation of the paper is as follows: In section \ref{sec:2} we review how to obtain the correlation function of links in $U(N)$ Chern-Simons theory. In section \ref{sec:3} we write down explicitly the $U(N)$ invariant of $(2,2m)$ torus link embedded inside three manifold $S^3$. In subsection \ref{subsec:4.1} we review the double scaling limit of $(2,2m)$ torus link invariant \cite{chakraborty2022large}.  The next subsection \ref{sec:5} contains the generalisation of this result for any $L$ component $(L\alpha,L\beta)$ torus link invariant. In section \ref{sec:torusmatrix} we discuss how one can express the $(L,L)$ torus link invariant in terms of correlation function involving $L$ Schur polynomials. An analytic expression for the leading contribution of any $(L,L)$ torus link invariant is obtained in the large $N$ limit using the saddle point approximation. Particularly we validate this result by tabulating the colored HOMFLY-PT polynomial of $(3,3)$ torus link. Subleading contributions to the link invariants have been discussed in section \ref{sec:subleading}. In that same section we also argue how the torus knot/link invariants with arbitrary representations can be expressed in terms of connected correlation functions of operators in $U(N)$ matrix model. Finally, in  the concluding section \ref{sec:conclu} we summarise our results and elaborate on the future outlook.

\section{Link invariants in \texorpdfstring{$U(N)$}{Lg} Chern-Simons theory}
\label{sec:2}
The three dimensional Chern-Simons theory  on a manifold $M$ corresponding to the gauge group $U(N)$ can be written as a sum of $SU(N)$ and $U(1)$ Chern-Simons actions \cite{borhade2004u}:
\begin{align}
    S&=\frac{k}{4\pi}\int_{M}\text{Tr}\bigg(A\wedge dA+\frac{2}{3}A\wedge A \wedge A\bigg)+\frac{k_{1}}{4\pi}\int_{M}\text{Tr}\bigg(B\wedge dB\bigg)\nonumber\\
    &\equiv \frac{k}{4\pi}\int_{M}d^{3}x ~\epsilon^{\mu \nu\rho}~\text{Tr}\left(A_{\mu}\partial_{\nu}A_{\rho}+\frac{2}{3}A_{\mu}A_{\nu}A_{\rho}\right)+\frac{k_{1}}{4\pi}\int_{M}d^{3}x~\epsilon^{\mu\nu\rho}~\text{Tr}\left(B_{\mu}\partial_{\nu}B_{\rho}\right)
\end{align}
where $A$, $B$ are the gauge connections for the gauge groups $SU(N)$, $U(1)$ respectively and $k$, $k_{1}$ are the respective coupling constants. Thus, the partition function of $U(N)$ Chern-Simons theory obtained, by integrating over gauge inequivalent connections, is a product of two partition functions 
\begin{align}
Z^{U(N)}[M]&=\int_{M}[\mathcal{D}A][\mathcal{D}B] e^{iS}\equiv Z^{SU(N)}[M]\times Z^{U(1)}[M].
\end{align}
As the action is explicitly metric independent, the corresponding partition function is a topological invariant of the three-manifold $M$. The other topological invariants associated with links can be described by expectation value of Wilson loop operators in Chern-Simons theory. For an oriented knot $\mathcal{K}$ in an irreducible representation $\mathcal{R}$, the Wilson loop operator is defined as
\begin{align}
W_{\mathcal{R}}(\mathcal{K})&= \bigg( \text{Tr}_{R}\mathbb{H}^{(A)}[\mathcal{K}]\bigg) \bigg(\text{Tr}_{n}\mathbb{H}^{(B)}[\mathcal{K}]\bigg)
\end{align}
where,
\begin{align}
\mathbb{H}^{(A)}[\mathcal{K}]&=\mathcal{P} \bigg[\exp \oint_{\mathcal{K}}A \bigg]~~\text{and}~~\mathbb{H}^{(B)}[\mathcal{K}]=\mathcal{P} \bigg[\exp \oint_{\mathcal{K}}B \bigg]
\end{align}
are the holonomies of $SU(N)$ and $U(1)$ gauge connection respectively. The symbol $\mathcal{P}$ denotes path ordering. Note that the representation $\mathcal{R}\in U(N)$ involves representation $R\in SU(N)$ and charge $n\in U(1)$. Given $(R,n)$, we can define $U(N)$ representation $\mathcal{R}$ in the following way:
\begin{align}
       n_{i}&=\begin{cases}
      l_{i}+s,~~~~i=1,...,N-1\\
      s~~~~,~~~~~~i=N\end{cases}\label{n_{i}def}
\end{align}
where $l_{i}$ ($n_{i}$) is the number of boxes in the $i^{th}$ row corresponding to the Young diagram of $SU(N)$ ($U(N)$) representation $R$ ($\mathcal{R}$) and $s$ is any integer. The total number of boxes of representation $\mathcal{R}$, known as the charge of $U(1)$ theory, is given by $n = \sum_{i=1}^{N}n_{i}$. The relation between $SU(N)$ Chern-Simons theory to $\mathfrak {su}(N )_k$ Wess-Zumino Witten conformal field theory implies that the $l_i$'s of the $SU(N)$ representations must obey $l_i \leq k$. Such representations are known as {\it integrable representations}. With suitable choice of $s$, we can shift the range of $n_i$ appropriately. 

If instead we have a link $\mathcal{L}$ made up of $L$ number of component knots $\mathcal{K}_{a}$ each carrying representation $\mathcal{R}_{a}$ of $U(N)$ gauge group, then the Wilson loop operator of $\mathcal{L}$ is
\begin{align}
W_{\mathcal{R}_{1},\cdots \mathcal{R}_{L}}[\mathcal{L}]&=\prod_{a=1}^{L} W_{\mathcal{R}_{a}}[\mathcal{K}_{a}]. 
\end{align} 
The correlation function of this topological operator yields the link invariant of $\mathcal{L}$ corresponding to $U(N)$ Chern-Simons theory
\begin{align}
\mathcal{W}^{\mathcal{L}}_{\mathcal{R}_1,...,\mathcal{R}_L}({M},k,k_{1})&=\langle {W}_{\mathcal{R}_1,...,\mathcal{R}_L}[\mathcal{L}] \rangle=\frac{1}{Z(M)}\int_{M} [\mathcal{D}A][\mathcal{D}B]\bigg(\prod_{a=1}^{L} {W}_{\mathcal{R}_a}[\mathcal{K}_{a}] \bigg)e^{iS}\nonumber\\
     &=\mathcal{W}^{\mathcal{L}}_{n^{(1)},\cdots,n^{(L)}}(M,k_{1})\mathcal{W}^{\mathcal{L}}_{R_{1},\cdots,R_{L}}(M,k).\label{U(N) invariant of CS ThEoRY}
 \end{align}
These link invariants can be computed using one of the two  framings:\\
$\bullet$ {\it standard framing} where the self-linking numbers of the component knots are zero, \\
$\bullet$ {\it vertical framing} where the self-linking number of any component knot is equal to its crossing number. 

In fact, the link invariants in standard framing are unchanged under all the three Reidemeister moves and are called {\it ambient isotopy invariants}. However the link invariants in vertical framing, referred to as {\it framed link invariants}, are unchanged only under the 
Reidemeister moves II and III. These framed link invariants are known in the knot theory literature as  {\it regular isotopy invariants}.

In this work, we will confine ourselves to  {framed links} [$\mathcal{L};f$] where the framing  $f=\{f_{1},\cdots, f_{L}\}$ is a set of $L$ integers, each element of the set denoting the self-linking number (or framing number) of the component knot. 

Let us consider the three manifold $M$ to be $S^3$. The $U(1)$ invariant of $[\mathcal{L};f]$ is given as 
\begin{align}
\mathcal{W}^{[\mathcal{L};f]}_{n^{(1)},\cdots,n^{(L)}}(S^{3},k_{
1})&=\exp\bigg(\frac{i\pi}{k_{1}}\sum_{a=1}^{L}f_{a}\big(n^{(a)}\big)^2\bigg)\exp\bigg(\frac{i\pi}{k_{1}}\sum_{a\ne b}^{L}(lk_{ab}) n^{(a)}n^{(b)}\bigg)\label{U(1)InvarianT}
\end{align} 
where $lk_{ab}$ denotes the linking number between the components $\mathcal{K}_{a}$ and $\mathcal{K}_{b}$. For the polynomial invariant of $U(N)$ Chern-Simons theory to be a function only of two variables
\begin{align}
    q=\exp\left( \frac{2\pi i}{k+N}\right)~\text{and}~v=q^{N},\label{VariableS}
\end{align}
in the $U(1)$ invariant (\ref{U(1)InvarianT}) we need to make the following replacement for the charge $n^{(a)}$ and the coupling constant $k_{1}$ \cite{borhade2004u, marino2001framed}  
\begin{align}
n^{(a)}\rightarrow \frac{n^{(a)}}{\sqrt{N}}~~,~~k_{1}\rightarrow k+N .
\end{align} 

We will elaborate the salient features of computing the $(L\alpha , L\beta)$ torus link invariants in the following section. 

\section{Torus link invariants  in \texorpdfstring{$U(N)$}{Lg} Chern-Simons theory}
\label{sec:3}
\begin{figure}
     \centering
     \begin{subfigure}[b]{0.40\textwidth}
         \centering
    \includegraphics[width=\textwidth]{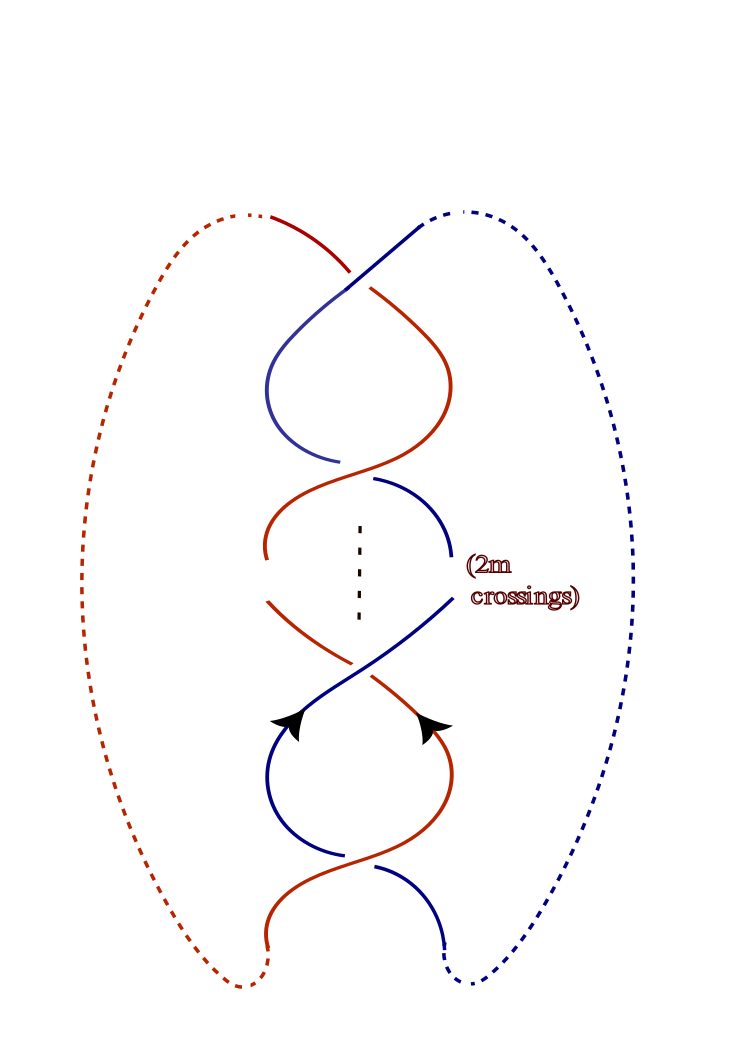}
         \caption{Right handed\\  (linking number +$m$)}
         \label{Right handed }
     \end{subfigure}
     \begin{subfigure}[b]{0.40\textwidth}
         \centering         \includegraphics[width=\textwidth]{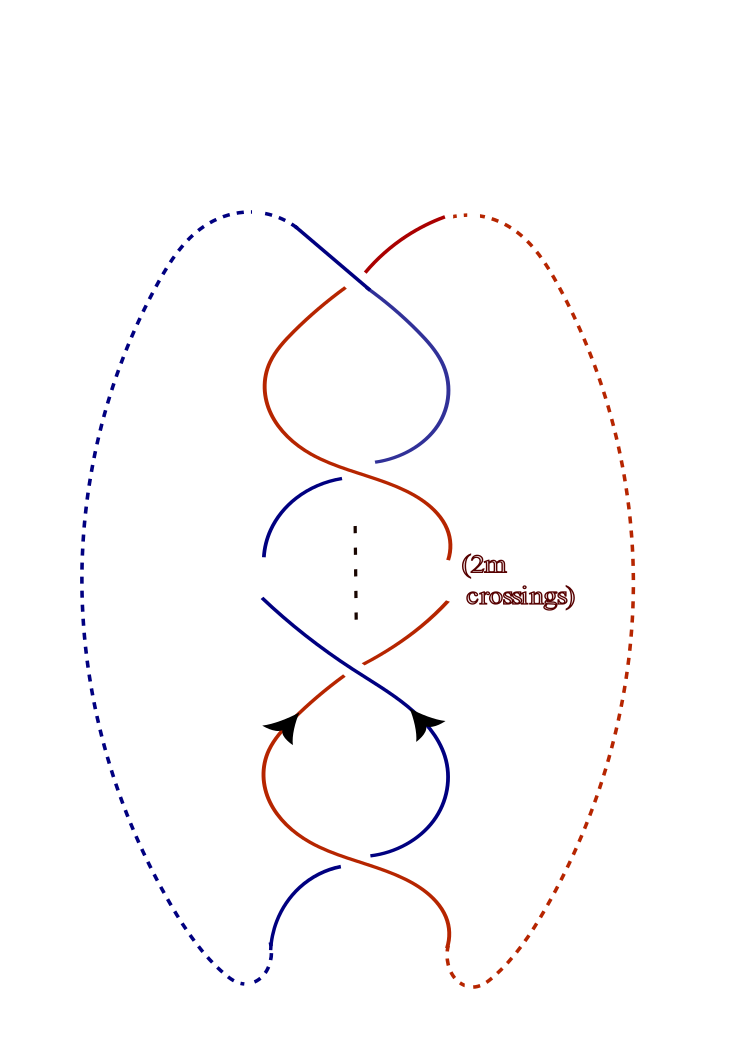}
         \caption{Left handed \\
         (linking number -$m$)}
         \label{Left handed}
     \end{subfigure}
     \caption{$(2,2m)$ torus link: closure of two parallely oriented strands with $2m$ crossings }
         \label{Link diagrams}
\end{figure}

Torus links with $L$ components are characterised by two integers $(L\alpha, L\beta)$ where $\alpha$ and $\beta$ are coprime to each other. These links can be obtained from the closure of a braid with $L\alpha$ strands. The braid word for a right handed $(L\alpha,L\beta)$ torus link, in terms of the braid generators $b_{i}$, is
\begin{align}
    \mathcal{B}(L\alpha,L\beta)&=\left(b_{1}b_{2}...b_{L\alpha-1}\right)^{L\beta}.\label{BraidWord}
\end{align}
The corresponding mirror image of this link is left handed and its braid word becomes $\left(b^{-1}_{1}b^{-1}_{2}...b^{-1}_{L\alpha-1}\right)^{L\beta}$. As an illustration, we show in figure (\ref{Link diagrams}) a two component ($L=2$) torus link (with $\alpha=1,~\beta=m$) and its mirror image , where $m$ is any positive integer.  The dotted lines in this figure indicate the closure of the two-strand braid.
For clarity, we now discuss the colored HOMFLY-PT polynomial for these $(2,2m)$ torus links before getting into arbitrary $L$ component torus links.
\subsection{\texorpdfstring{$(2,2m)$}{Lg} torus link invariant}
Given a braid word (\ref{BraidWord}), there is a systematic procedure of directly writing down the invariants of links that are obtained from the closure of such  braids \cite{RamaDevi:1992np}. Basically, we use the basis where the braiding generators are diagonal. As any two consecutive generators $b_i$ and $b_{i\pm 1}$ do not commute, we need to perform a suitable unitary transformation to go from one basis to another. The invariant of the right-handed $(2,2m)$ torus link (refer to Fig. \ref{Right handed }), closure of a simple two-strand braid with braid word $b_{1}^{2m}$, involves the eigenvalues only of the braiding generator $b_{1}$.
In vertical framing, the eigenvalues for the right handed braiding generator $b_1$ on two parallely oriented strands  carrying representations ${R}_{1}$ and ${R}_{2}$ of $SU(N)$ gauge group is given by \cite{kaul2001three}
\begin{align}
    \lambda_{R_t}^{(+)}(R_1,R_2)&=(-1)^{\epsilon_{R_1}+\epsilon_{R_{2}}-\epsilon_{R_t}}q^{-\big(\frac{C_{R_{1}}+C_{R_2}}{2}\big)+\frac{C_{R_t}}{2}},~~~~~R_t\in R_1\otimes R_2.\label{EigenValuE}
\end{align}
where $\epsilon_{R_1},\epsilon_{R_2},\epsilon_{R_{t}}=\pm 1$,  $q$ is defined in equation (\ref{VariableS}) and $C_{R_{a}}$, the quadratic Casimir of representation $R_{a}$ is defined as
\begin{align}
C_{R_{a}}&=-\frac{\big(l^{(a)}\big)^2}{2N}+\frac{1}{2}\sum_{j}l_{j}^{(a)}\big(l_{j}^{(a)}-2j+N+1\big).
\end{align}
Here $l_{i}^{(a)}$ denotes the number of boxes in the $i$th row of representation $R_{a}$ and $l^{(a)}=\sum_{i}l_{i}^{(a)}$. Using the above braiding eigenvalue (\ref{EigenValuE}), we can write down the colored HOMFLY-PT polynomial of $(2,2m)$ torus link with framing $f=\{f_{1},f_{2}\}$ of the component knots and linking number      $m$ \cite{borhade2004u}:
\begin{align}
\mathcal{W}_{R_{1},R_{2}}^{[(2,2m);f]}(S^3,k)&=q^{f_{1}C_{R_{1}}+f_{2}C_{R_2}}\sum_{R_{t}\in R_{1}\otimes R_2}\big(\text{dim}_{q}R_{t}\big)\bigg(\lambda^{(+)}_{{R}_{t}}\bigg)^{2m}\nonumber\\
&=q^{f_{1}C_{R_{1}}+f_{2}C_{R_2}}\sum_{R_t\in R_{1}\otimes R_{2}}(\text{dim}_{q}R_t)\bigg(q^{-\frac{C_{R_1}+C_{R_2}}{2}+\frac{C_{R_t}}{2}}\bigg)^{2m},\label{SU(N)2,2mFramed}
\end{align}
where the $q$-dimension of a representation is defined as
\begin{align}
\text{dim}_{q}R_{a}&=\prod_{1\le i<j \le N}\frac{\big[l_{i}^{(a)}-l_{j}^{(a)}+j-i\big]_{q}}{[j-i]_{q}}~~\text{and}~~[x]_{q}=\frac{q^{\frac{x}{2}}-q^{-\frac{x}{2}}}{q^{\frac{1}{2}}-q^{-\frac{1}{2}}}.
\end{align}
The $U(N)$ invariant of framed $(2,2m)$ torus link can be written using (\ref{U(N) invariant of CS ThEoRY}, \ref{U(1)InvarianT}, \ref{SU(N)2,2mFramed}): 
\begin{align}
\mathcal{W}_{\mathcal{R}_{1},\mathcal{R}_{2}}^{[(2,2m);f]}(S^3,k)&=\mathcal{W}_{n^{(1)},n^{(2)}}^{[(2,2m);f]}(S^3,k)\times\mathcal{W}_{R_{1},R_{2}}^{[(2,2m);f]}(S^3,k)\nonumber\\
&=\exp\bigg(\frac{i\pi}{k+N}\sum_{a=1}^{2}\frac{f_{a}\big(n^{(a)}\big)^2}{N}\bigg)\exp\bigg(\frac{i\pi}{k+N}\frac{2(+m)n^{(1)}n^{(2)}}{N}\bigg)\nonumber\\
&~~~\times q^{f_{1}C_{\mathcal{R}_{1}}+f_{2}C_{\mathcal{R}_2}}\sum_{\mathcal{R}_t\in \mathcal{R}_{1}\otimes \mathcal{R}_{2}}(\text{dim}_{q}\mathcal{R}_t)\bigg(q^{-\frac{C_{\mathcal{R}_1}+C_{\mathcal{R}_2}}{2}+\frac{C_{\mathcal{R}_t}}{2}}\bigg)^{2m}.
\end{align}
Note that, in writing the above expression we have used the fact that the quadratic Casimir and $q$-dimension of a $U(N)$ representation $\mathcal{R}_{a}$ is same as that of the quadratic Casimir and $q$-dimension of $SU(N)$ representation $R_{a}$:
\begin{align}
    C_{\mathcal{R}_{a}} & = -\frac{\big(n^{(a)}\big)^2}{2N} + \frac{1}{2}\sum_{i=1}^{N}n_{i}^{(a)}\big(n_{i}^{(a)} - 2i + N + 1 \big) = C_{R_{a}}.\\ \text{dim}_{q}&\mathcal{R}_{a}=\prod_{1\le i<j\le N}\frac{\big[n_{i}^{(a)}-n_{j}^{(a)}+j-i\big]_{q}}{\big[j-i\big]_{q}}=\text{dim}_{q}R_{a}.
\end{align}
We introduce $\kappa_{\mathcal{R}_{a}}$ in terms of the number of boxes $n_{i}^{(a)}$ in the $i$th row of the Young diagram of representation $\mathcal{R}_{a}$:
\be\label{eq:kappadef}
\kappa_{\mathcal{R}_{a}} = \frac{1}{2}\sum_{i=1}^{N}n_{i}^{(a)}\big(n_{i}^{(a)} - 2i + N + 1 \big).
\ee
In terms of these variables, we express the $U(N)$ invariant of the framed $(2,2m)$ torus link as
  \begin{align}
\mathcal{W}_{\mathcal{R}_{1},\mathcal{R}_{2}}^{[(2,2m);f]}(S^3,k)&=q^{f_{1}\kappa_{\mathcal{R}_{1}}+f_{2}\kappa_{\mathcal{R}_{2}}}\sum_{\mathcal{R}_{t}}(\text{dim}_{q}\mathcal{R}_t)\bigg(q^{-\kappa_{\mathcal{R}_{1}}-\kappa_{\mathcal{R}_{2}}+\kappa_{\mathcal{R}_{t}}}\bigg)^{m}.\label{U(N)invAriNantTof2,2mToruSliNK}
  \end{align}
   Hereafter we refer $\kappa_{\mathcal{R}_{a}}$ to be the quadratic Casimir of $U(N)$ representation $\mathcal{R}_{a}$. The invariants for arbitrary $L$-component torus links can be compactly written using these quadratic casimirs $\kappa_{\mathcal{R}_{a}}$ and $q$-dimension of representations \cite{lin2010hecke}. We will briefly present the explicit form of these link invariants in the following subsection.
\subsection{\texorpdfstring{$(L\alpha , L \beta)$}{Lg} torus link invariant}
   For a general two-component right handed torus link $(2\alpha, 2 \beta)$ carrying $U(N)$ representations $\mathcal{R}_{1}$ and $\mathcal{R}_{2}$ on the  component knots, the polynomial invariant is \cite{lin2010hecke}
\begin{align}
 \mathcal{W}^{(2\alpha,2\beta)}_{\mathcal{R}_{1},\mathcal{R}_{2}}(S^3,k)&=\prod_{i=1}^{2}\left(q^{-\frac{\alpha \beta}{2}\varkappa_{\mathcal{R}_{i}}}~v^{-\frac{\beta(\alpha-1)}{2}n^{(i)}}\right)\sum_{\mathcal{R}_{t}\vdash \alpha(n^{(1)}+n^{(2)})}\zeta^{\mathcal{R}_{t}}_{\mathcal{R}_{1},\mathcal{R}_{2}}q^{\frac{\beta}{2\alpha}\varkappa_{\mathcal{R}_{t}}}s^{\star}_{\mathcal{R}_{t}}(q,v)\label{2Alp,2BetAInvAriANt}
 \end{align}
 where
 \begin{align}
 \zeta^{\mathcal{R}_{t}}_{\mathcal{R}_{1},\mathcal{R}_{2}}&=\sum_{\Lambda^{(1)}\vdash n^{(1)}}\frac{|c_{\Lambda^{(1)}}|}{(n^{(1)})!}\chi_{\mathcal{R}_{1}}(c_{\Lambda^{(1)}})\sum_{\Lambda^{(2)}\vdash n^{(2)}}\frac{|c_{\Lambda^{(2)}}|}{(n^{(2)})!}\chi_{\mathcal{R}_{2}}(c_{\Lambda^{(2)}})\chi_{\mathcal{R}_{t}}(c_{(\Lambda^{(1)}+\Lambda^{(2)})_{(\alpha)}}),\nonumber\\
s_{\mathcal{R}_{t}}^{\star}(q,&v)=\sum_{\Lambda\vdash n^{(t)}}\frac{|c_{\Lambda}|}{(n^{(t)})!}\chi_{\mathcal{R}_{t}}(c_{\Lambda})\prod_{i=1}^{\ell(\Lambda)}\frac{v^{\Lambda_{i}/2}-v^{-\Lambda_i/2}}{q^{\Lambda_i/2}-q^{-\Lambda_i/2}}\equiv\text{dim}_{q}\mathcal{R}_{t},~~v=q^{N}.
\end{align}
Here $\varkappa_{\mathcal{R}}$ in terms of $\kappa_{\mathcal{R}}$ (\ref{eq:kappadef}) can be written as 
\begin{align}
    \varkappa_{\mathcal{R}}=\sum_{i}n_i(n_i-2i+1)=2\kappa_{\mathcal{R}}-N\sum_{i}n_{i}.
\end{align}
The notation $\Lambda^{(a)}\vdash n^{(a)}$ means that  $\Lambda^{(a)}\equiv\left(\Lambda^{(a)}_{1},\Lambda^{(a)}_{2},\Lambda^{(a)}_{3},... \right)$ is a partition, in other words a sequence of positive integers with $\Lambda_{1}^{(a)}\ge \Lambda_{2}^{(a)}\ge \Lambda_{3}^{(a)}\ge \cdots$, satisfying 
\begin{align}
&\sum_{i=1}^{\ell(\Lambda^{(a)})}\Lambda^{(a)}_{i}=n^{(a)},
\end{align}
where $\ell(\Lambda^{(a)})$ is the highest value of $i$ for which $\Lambda^{(a)}_{i}$ is non zero:
\begin{align}
    \ell(\Lambda^{(a)})&=\mathrm{max}\left\{i \vert \Lambda^{(a)}_{i}>0 \right\}.
\end{align}
 The conjugacy class associated to the partition $\Lambda^{(a)}$ is denoted as $ c_{\Lambda^{(a)}}$ and it consists of one $\Lambda^{(a)}_{1}$-cycle, one $\Lambda^{(a)}_{2}$-cycle, one $\Lambda^{(a)}_{3}$-cycle and so on. $|c_{\Lambda^{(a)}}|$ denotes the total number of elements belonging to the conjugacy class $c_{\Lambda^{(a)}}$, having the same cycle structure. $\chi_{\mathcal{R}_{a}}(c_{\Lambda^{(a)}})$ is the corresponding character labelled by representation $\mathcal{R}_{a}$. The sum of two partitions is the sum of each of the elements of the respective partitions
\begin{align}
    \Lambda^{(1)}+\Lambda^{(2)}\equiv \left( \Lambda^{(1)}_{1}+\Lambda^{(2)}_{1},\Lambda^{(1)}_{2}+\Lambda^{(2)}_{2},\Lambda^{(1)}_{3}+\Lambda^{(2)}_{3},\cdots\right),
\end{align}
and $(\Lambda)_{(\alpha)}$ implies the partition $(\alpha\Lambda_{1},\alpha \Lambda_{2},\alpha\Lambda_{3},...)$.

Rewriting the torus link invariant (\ref{2Alp,2BetAInvAriANt}) in terms of variable $\kappa_{\mathcal{R}}$ we get:
\begin{align}
\mathcal{W}^{(2\alpha,2\beta)}_{\mathcal{R}_{1},\mathcal{R}_{2}}(S^3,k)&=q^{-{\alpha\beta}\big(\kappa_{\mathcal{R}_{1}}+\kappa_{\mathcal{R}_{2}}\big)}\sum_{\mathcal{R}_{t}\vdash \alpha(n^{(1)}+n^{(2)})} \zeta^{\mathcal{R}_{t}}_{\mathcal{R}_{1},\mathcal{R}_{2}}q^{\frac{\beta}{\alpha}\kappa_{\mathcal{R}_{t}}} s^{\star}_{\mathcal{R}_{t}}(q,v).
\end{align}
If the component knots are having $f_{1}$ and $f_{2}$ self-linking numbers, which we denote as $f=\{f_{1},f_{2}\}$, then the polynomial invariant will have  an additional framing factor
\begin{align}
\mathcal{W}^{[(2\alpha,2\beta);f]}_{\mathcal{R}_{1},\mathcal{R}_{2}}(S^3,k)&=q^{f_{1}\kappa_{\mathcal{R}_{1}}+f_{2}\kappa_{\mathcal{R}_{2}}}\mathcal{W}^{(2\alpha,2\beta)}_{\mathcal{R}_{1},\mathcal{R}_{2}}(S^3,k).\label{(2alpha,2BetAInv)}
\end{align}
Now, we can generalise the colored HOMFLY-PT polynomial for an arbitrary $L$ component torus link $(L\alpha ,L\beta)$, with self-linking number $f=\{f_{1},...f_{L}\}$ of the component knots, as 
\begin{align}
\mathcal{W}^{[(L\alpha,L\beta);f]}_{\mathcal{R}_{1},...,\mathcal{R}_{L}}(S^3,k)=\bigg(\prod_{a=1}^{L}&q^{(f_{a}-\alpha \beta)\kappa_{\mathcal{R}_{a}}}\bigg)\sum_{\mathcal{R}_{t}\vdash \alpha(n^{(1)}+\cdots+n^{(L)})} \zeta^{\mathcal{R}_{t}}_{\mathcal{R}_{1},...,\mathcal{R}_{L}}q^{\frac{\beta}{\alpha}\kappa_{\mathcal{R}_{t}}} s^{\star}_{\mathcal{R}_{t}}(q,v)\label{LAlp,LBeTa}
\end{align}
where
\begin{align}
\zeta^{\mathcal{R}_{t}}_{\mathcal{R}_{1},...,\mathcal{R}_{L}}=\prod_{a=1}^{L}\left(\sum_{\Lambda^{(a)}\vdash n^{(a)}}\frac{|c_{\Lambda^{(a)}}|}{(n^{(a)})!}\chi_{\mathcal{R}_{a}}(c_{\Lambda^{(a)}})\right)\chi_{\mathcal{R}_{t}}(c_{(\Lambda^{(1)}+\cdots+\Lambda^{(L)})_{(\alpha)}}).
\end{align}
In Ref. \cite{chakraborty2022large}, two of the authors have analysed the double scaling limit (\ref{eq:dslimit1},\ref{eq:dslimit2}) of the two-component $(2,2m)$ torus link invariants. Particularly, these invariants were shown to be expressed in terms of $(2,2)$ torus link invariants. We believe that this must be generalisable for an arbitrary $L$-component torus link $(L\alpha, L\beta)$. In the following section we present our result after briefly reviewing the works on 
$(2,2m)$ torus links.
\section{Torus link invariants in the double scaling limit of \texorpdfstring{$U(N)$}{Lg} Chern-Simons theory}
\label{sec: 4}
As discussed in the introduction, the double scaling limit (\ref{eq:dslimit1}) of $U(N)$ Chern-Simons theory involves the Chern-Simons coupling $k \rightarrow \infty$. The range of the $n^{(a)}_i$'s (\ref{n_{i}def}) for the integrable representations $\mathcal{R}_{a}$ at large $k$  can be chosen as 
\begin{equation}\label{eq:integrabilitycond}
-\frac{k}{2} \leq n_N^{(a)}\leq \cdots \leq n_1^{(a)}\leq \frac{k}{2}.
\end{equation}
For the representation $\mathcal{R}_{a}$ we define a set of $N$ variables $\{\theta_{1}^{(a)},...,\theta_{N}^{(a)}\}$: 
\begin{align}
    \theta_{i}^{(a)}&=\frac{2\pi}{N+k}\bigg(h_{i}^{(a)}-\frac{N-1}{2}\bigg), \label{eq:theta-a}\\
    \text{where,} \quad h_{i}^{(a)} & = n_{i}^{(a)}+N-i.\label{eq:h-a}
\end{align}
It is easy to check that in the double scaling limit the variables $\theta_i^{(a)}$ range from $-\pi$ to $\pi$. The quadratic Casimir of representation $\mathcal{R}_{a}$ (\ref{eq:kappadef}) can be expressed in terms of these variables as
\begin{align}
\kappa_{\mathcal{R}_a} & =  \frac{(N+k)^2}{8\pi^2}\sum_{i}{\theta_{i}^{(a)}}^2 -\frac{N(N^2-1)}{24}. \label{QuadraticCasimirOfU(N)inTermsofTheta}
\end{align}
In the large $N$ limit, the set of discrete variables $\theta_i$ becomes a continuous function 
\begin{align}
   \theta^{(a)}_{i}\rightarrow \theta^{(a)}(x)&~~~~\text{where}~~ x = \frac{i}{N} \in [0,1].\label{ContiNuousVar}
\end{align}
All the discrete summations over $i$ are then replaced by integration over $x$
\begin{equation}
    \frac{1}{N}\sum_{i} = \int_0^1 dx.\label{SumtoInt}
\end{equation}
Moreover the distribution of $\theta_i$, in the large $N$ limit is captured by a distribution function
\begin{equation}
    \sigma(\theta) = \frac{1}{N} \sum_{i=1}^N \delta(\theta - \theta_i) = \left | \frac{\partial x}{\partial \theta} \right |,\label{DisTributioNofeigv}
\end{equation}
such that $\int d\theta \sigma(\theta) =1$. Hence, for any arbitrary functions of $\theta_i$ we can write
\begin{eqnarray}
   \frac{1}{N} \sum_i f(\theta_i) \rightarrow  \int_0^1 dx f(\theta(x))= \int d\theta \ \sigma(\theta) f(\theta). 
\end{eqnarray}
Now we will incorporate the continuum limit discussed above for the $(2,2m)$ torus link invariant.
\subsection{\texorpdfstring{($2,2m$)}{Lg} torus link invariant in the double scaling limit}
\label{subsec:4.1}
The invariant $\mathcal{W}_{\mathcal{R}_{1},\mathcal{R}_{2}}^{[(2,2m);f]}(S^3,k)$ (\ref{U(N)invAriNantTof2,2mToruSliNK}) can be expressed using the $\{\theta_{i}\}$ variables (\ref{eq:theta-a}) as
\begin{equation}
\begin{aligned}
 \mathcal{W}_{\mathcal{R}_{1},\mathcal{R}_{2}}^{[(2,2m);f]}(S^3&,k)  =\exp\bigg[\frac{i(N+k)}{4\pi}\bigg
 ((f_1-m)\sum_{i}{\theta_{i}^{(1)}}^2+(f_2-m)\sum_{i}{\theta_{i}^{(2)}}^{2}\bigg)\bigg] \\
 \times \exp \bigg[- &\frac{i\pi(N^3-N)}{12(N+k)}(f_1+f_2-m)\bigg] \times \sum_{\{\theta^{(t)}_{i}\}} \frac{\textstyle \exp\bigg[\frac{1}{2}\sum_{i\ne j}^{N}\log\left|\sin\left(\frac{\theta^{(t)}_{i}-\theta^{(t)}_{j}}{2}\right)\right|\bigg]}{\textstyle \exp\bigg[\frac{1}{2}\sum_{i\ne j}^{N}\log~\left|\sin\left(\frac{\pi(j-i)}{N+k}\right)\right|\bigg]}\\
 & \times \exp\bigg[\frac{im(N+k)}{4\pi}\sum_{i}{\theta^{(t)}_{i}}^2\bigg].
 \end{aligned}
 \end{equation}
Here $\theta_i^{(1)}$, $\theta_i^{(2)}$ and $\theta_i^{(t)}$ are the set of variables corresponding to representations $\mathcal{R}_1$, $\mathcal{R}_2$ and $\mathcal{R}_t$ respectively. In the double scaling limit with appropriate replacement of the discrete variables in terms of the continuous functions (\ref{ContiNuousVar}, \ref{SumtoInt}, \ref{DisTributioNofeigv}) the above invariant becomes
\begin{align}
{\mathcal{W}}^{[(2,2m);f]}&(S^{3},\lambda) = {\mathcal{G}}^{(f)}(m,\lambda,\sigma_1,\sigma_2) \mathcal{F}(m,\lambda)
\end{align}
where
{\begin{equation}
    \begin{aligned}
    {\mathcal{G}}^{(f)}(m,&\lambda,\sigma_1,\sigma_2)=\exp\bigg[\frac{iN^{2}}{4\pi \lambda}\bigg(\int d\theta \theta^2 \left(\left(f_1 -m\right){\sigma_{1}(\theta)}+\left( f_2 -m\right){\sigma_{2}(\theta)}\right)\bigg)\bigg]\\
  &~~~~~~\times \exp\bigg[-\frac{i\pi \lambda N^2}{12}(f_1+f_2-m)- \frac{N^2}{2}\int dx \dashint dy \log \left|\sin\left(\pi \lambda(y-x)\right)\right|\bigg]
    \end{aligned}
\end{equation}}
is dependent on framing $f$ and
{ \begin{equation}
    \begin{aligned}
    \mathcal{F}(m,\lambda) &=\int [\mathcal{D}\theta^{(t)}(x)] \exp\bigg[\frac{N^2}{2}\int dx \dashint dy~  \log~\bigg|\sin\left(\frac{\theta^{(t)}(x)-\theta^{(t)}(y)}{2}\right)\bigg|\bigg]\\
&~~~~~~~~~~~~~~~~\times\exp\bigg[\frac{imN^2}{4\pi \lambda}\int dx ~\theta^{(t)}(x)^2\bigg]
\end{aligned}
\end{equation}}  
is independent of framing.
Observe that the function ${\mathcal{G}^{(f)}}(m,\lambda,\sigma_1,\sigma_2)$ depends only on the given representations $\mathcal{R}_1$ and $\mathcal{R}_2$ and is independent of $\mathcal{R}_t$. However, $\mathcal{F}(m,\lambda)$ involves the $\theta^{(t)}$ integration which is constrained, since it runs only over the irreducible representations  $ \mathcal{R}_{t}$ appearing in the tensor product of $\mathcal{R}_1$ and $\mathcal{R}_2$. Therefore, this function $\mathcal{F}(m,\lambda)$ is difficult to evaluate even in the large $N$ limit. Note that, the space of functional integration over $\theta^{(t)}(x)$ in $\mathcal{F}(m,\lambda)$ remains the same for any value of $m$.  As a result, the function $\mathcal{F}(m,\lambda)$ has the following property
\begin{equation}
    \mathcal{F}(m,\lambda) = \mathcal{F}(1,\lambda/m). \label{SymmetryofF(m,lambda)}
\end{equation}
With $m=1$, $ \mathcal{W}_{\mathcal{R}_{1}, \mathcal{R}_{2}}^{[(2,2m);f]}(S^3,k)$ becomes the invariant of framed Hopf link. Therefore, by exploiting the property (\ref{SymmetryofF(m,lambda)}), we can see 
that a framed $(2,2m)$ torus link invariant can be expressed in terms of a framed Hopf link invariant as follows:
\begin{align}
{\mathcal{W}}_{\sigma_{1},\sigma_{2}}^{[(2,2m);f]}(S^3,\lambda)&=\bigg(\frac{\mathcal{G}^{(f)}(m,\lambda,\sigma_1,\sigma_2)}{\mathcal{G}^{(f)}(1,\lambda/m,\sigma_1,\sigma_2)}\bigg)~\mathcal{W}_{\sigma_{1},\sigma_{2}}^{[(2,2);f]}(S^3,\lambda /m).\label{2,2m framed link invariant in terms of 2,2 framed link inv}
\end{align}
That is, we first find the distribution functions $\sigma_1, \sigma_2$ corresponding to the representations
$\mathcal{R}_1$, $\mathcal{R}_2$ placed on the two component knots. After which we can write the Hopf link invariant as a function of $\sigma_{1}$, $\sigma_2$ and $\lambda$. 
Then the link invariant for $(2,2m)$ torus link as a function of $\lambda$ becomes proportional to the Hopf link invariant with $\lambda$ replaced by $ \lambda/m$ but importantly, keeping $\sigma_{1}$ and $\sigma_{2}$ fixed while performing such a replacement.

In the following subsection we generalise (\ref{2,2m framed link invariant in terms of 2,2 framed link inv}) for any $L$ component torus link $(L\alpha,L\beta)$ embedded inside $S^3$.

\subsection{\texorpdfstring{$(L\alpha , L\beta)$}{Lg} torus link invariant in the double scaling limit}
\label{sec:5}
Let us first focus on arbitrary 2 component torus links $(2\alpha,2\beta)$. Using the expression of $\kappa_{\mathcal{R}_{a}}$ in terms of the set of $N$ variables $\theta_{i}^{(a)}$ (\ref{QuadraticCasimirOfU(N)inTermsofTheta}) we can express the $(2\alpha,2\beta)$ torus link invariant (\ref{(2alpha,2BetAInv)}) as
{\begin{equation}
\begin{aligned}
&\mathcal{W}^{[(2\alpha,2\beta);f]}_{\mathcal{R}_{1},\mathcal{R}_{2}}(S^3,k)=\exp\bigg[\frac{i(N+{k})}{4\pi}\bigg((f_1-\alpha\beta)\sum_{i}{\theta_{i}^{{(1)}}}^2+(f_2-\alpha\beta)\sum_{i}{\theta_{i}^{{(2)}}}^2\bigg)\bigg]\\
    &~~~~~~~~~~~~~~~~~~~~~~~~~~~~~~~~~\times\exp\bigg[-\frac{i\pi(N^3-N)}{12(N+{k})}(f_1+f_2-2\alpha\beta+\frac{\beta}{\alpha})\bigg]\\
    &~~~~~\times\sum_{\{\theta^{(t)}_{i}\}}\zeta^{\theta^{(t)}}_{\theta^{{(1)}},\theta^{{(2)}}} \exp\bigg[\frac{i\beta(N+{k})}{4\pi \alpha}\sum_{i}{\theta^{(t)}_{i}}^2\bigg]\frac{\textstyle \exp\bigg[\frac{1}{2}\sum_{i\ne j}^{N}\log\left|\sin\left(\frac{\theta^{(t)}_{i}-\theta^{(t)}_{j}}{2}\right)\right|\bigg]}{\textstyle \exp\bigg[\frac{1}{2}\sum_{i\ne j}^{N}\log~\left|\sin\left(\frac{\pi(j-i)}{N+k}\right)\right|\bigg]}.\label{HopfLinkInv2alpha2beta}
\end{aligned}
\end{equation}}
We now take the double scaling limit with the appropriate replacement of the discrete variables in terms of the continuous functions (\ref{ContiNuousVar}, \ref{SumtoInt}, \ref{DisTributioNofeigv}):
{
    \begin{align}
& \mathcal{W}^{[(2\alpha,2\beta);f]}_{{\sigma}_{1},{\sigma}_{2}}(S^3,\lambda)=\exp\bigg[\frac{iN^2}{4\pi \lambda}\bigg(\int d\theta \theta^{2}\big
 ((f_1-\alpha\beta) \sigma_{1}(\theta)+(f_2-\alpha\beta)\sigma_{2}(\theta)\big)\bigg)\bigg]\nonumber\\
 &~~~~~~~~~~~~~~~~~~~~~~~~~~~~~~~~~~~~~~~~\times\exp\bigg[-\frac{i\pi \lambda N^2}{12}(f_1+f_2-2\alpha\beta+\frac{\beta}{\alpha})\bigg]\\
    \times& \int [\mathcal{D}\theta^{(t)}]\zeta^{\theta^{(t)}}_{\theta^{(1)},\theta^{(2)}} \exp\bigg[\frac{i\beta N^2}{4\pi \alpha \lambda}\int d\theta \theta^{2}\sigma_{t}(\theta)\bigg] \frac{\textstyle \exp\bigg[\frac{N^2}{2}\int dx \dashint dy\log\bigg|\sin\left(\frac{\theta^{(t)}(x)-\theta^{(t)}(y)}{2}\right)\bigg|\bigg]}{\textstyle \exp\bigg[\frac{N^2}{2}\int dx\dashint dy\log\big|\sin\left({\pi \lambda(y-x)}\right)\big|\bigg]}\nonumber\\
&~~~~~~~\equiv{\mathcal{G}}^{(f)}(\alpha,\beta,\lambda,\sigma_{1},\sigma_{2}){\mathcal{F}}\bigg(\frac{\beta}{\alpha},\lambda \bigg),
\end{align}    
where
\begin{equation}
\begin{aligned}
  &  {\mathcal{G}}^{(f)}(\alpha,\beta,\lambda,\sigma_{1},\sigma_{2})=\exp\bigg[\frac{iN^2}{4\pi \lambda}\bigg(\int d\theta \theta^{2}\big
 ((f_1-\alpha\beta) \sigma_{1}(\theta)+(f_2-\alpha\beta)\sigma_{2}(\theta)\big)\bigg)\bigg]\\
 &~~\times\exp\bigg[-\frac{i\pi \lambda N^2}{12}\left(f_1+f_2-2\alpha\beta+\frac{\beta}{\alpha}\right)-\frac{N^2}{2}\int dx\dashint dy~ \log \big|\sin \left(\pi \lambda (y-x)\right)\big|\bigg],
 \end{aligned}
 \end{equation}
 and
 \begin{equation}
 \begin{aligned}
 &{\mathcal{F}}\bigg(\frac{\beta}{\alpha},\lambda \bigg) =\int [\mathcal{D}\theta^{(t)}]\zeta^{\theta^{(t)}}_{\theta^{(1)},\theta^{(2)}} \exp\bigg[\frac{N^2}{2}\int dx\dashint~dy~ \log~ \bigg|\sin\left( \frac{\theta^{(t)}(x)-\theta^{(t)}(y)}{2}\right)\bigg|\bigg]\\
&~~~~~~~~~~~~~~~~~~~~~~~~~~~~~~~~~~~~~~~~\times \exp\bigg[\frac{i\beta N^2}{4\pi \alpha \lambda}\int d\theta \theta^{2}\sigma_{t}(\theta)\bigg].
\end{aligned}
\end{equation}
When $\alpha=\beta=1$, we get the double scaling limit of Hopf link invariant.
We exploit the symmetry of ${\mathcal{F}}\bigg(\frac{\beta}{\alpha},\lambda\bigg)$ under
\begin{align}
    \frac{\beta}{\alpha}\rightarrow 1 ~~\text{and}~~\lambda\rightarrow \frac{\lambda}{\beta/\alpha}
\end{align} 
to write down the invariant of any 2 component torus link in terms of Hopf link invariant as
\begin{align}
\mathcal{W}^{[(2\alpha,2\beta);f]}_{{\sigma}_{1},{\sigma}_{2}}&(S^3,\lambda)=\frac{{\widetilde{\mathcal{G}}}^{(f)}(\alpha,\beta,\lambda,\sigma_{1},\sigma_{2})}{\widetilde{\mathcal{G}}^{(f)}(1,1,\frac{\lambda}{\beta/\alpha},\sigma_{1},\sigma_{2})}\mathcal{W}^{[(2,2);f]}_{{\sigma}_{1},{\sigma}_{2}}\bigg(S^3,\frac{\lambda}{\beta/\alpha}\bigg).
\end{align}
Similar exercise  for any arbitrary $L$ component torus link (\ref{LAlp,LBeTa}) is straightforward and we find that $(L\alpha,L\beta)$ torus link invariant can be expressed in terms of $(L,L)$ torus link invariant  as follows:
\begin{align}
\mathcal{W}^{[(L\alpha,L\beta);\{f_1,...,f_L\}]}_{{\sigma}_{1},...,{\sigma}_{L}}&(S^3,\lambda)=\frac{\widetilde{\mathcal{G}}^{(f)}(\alpha,\beta,\lambda,\sigma_{1},...,\sigma_{L})}{\widetilde{\mathcal{G}}^{(f)}(1,1,\frac{\lambda}{\beta/\alpha},\sigma_{1},...,\sigma_{L})}\mathcal{W}^{[(L,L);\{f_{1},...,f_{L}\}]}_{{\sigma}_{1},...{\sigma}_{L}}\bigg(S^3,\frac{\lambda}{\beta/\alpha}\bigg).\label{LAlphLbetaL,L}
\end{align}
Thus, given a $(L,L)$ torus link invariant in variable $\lambda$ with  $\sigma_{1},...,\sigma_{L}$ being the eigenvalue densities of the component knots, the  $(L\alpha,L\beta)$ torus link invariant (carrying the same eigenvalue densities) in variable $\lambda$ becomes  proportional to the $(L,L)$ torus link invariant with  $\lambda$  replaced by $\frac{\lambda}{\beta/\alpha}$ while keeping $\sigma_{1},...,\sigma_{L}$ unaltered.

 We can explicitly work out the $(L,L)$ link invariants (\ref{LAlp,LBeTa}) with components carrying different representations whose Young diagrams have small number of boxes. For these representations, we can infer the leading and subleading terms for a general $(L\alpha,L\beta)$ link invariant  in the double scaling limit (\ref{LAlphLbetaL,L}). Extracting such leading and subleading contributions appear to be practically difficult for representations with large number of boxes. In Ref. \cite{chakraborty2022large}, large $N$ contributions for the two component Hopf link invariant was determined by mapping it to a one dimensional fluid equation, where the initial and final fluid densities correspond to two of the representations placed on the two component knots. However, such a procedure is not generalisable for a $(L,L)$ torus link with $L>2$.
 Matrix model method appears to be an efficient approach to handle such difficulties. 
  In the following section, we will briefly review $U(N)$ matrix model and the computation of the correlators corresponding to the torus link invariants in the double scaling limit.
\section{Torus link invariants and matrix models at large \texorpdfstring{$N$}{Lg}}
\label{sec:torusmatrix}
 
Study of the partition function and the correlators in $U(N)$ matrix model at large $N$ will allow us to deduce the leading and subleading contributions of torus link invariants. Within the matrix model approach, we  obtain analytic results for these contributions at the following two instances:
\begin{itemize}
    \item representations having small number of boxes (small  as compared to $N,k \rightarrow \infty$), these are validated  with those obtained from explicit  colored HOMLFLY-PT polynomials.
 \item  large symmetric representations placed on the component knots.
\end{itemize}

We now focus on the invariant of a framed Hopf link, embedded on a three-manifold $S^3/\mathbb{Z}_p$, which can equivalently be written in terms of the modular transformation matrices ($\mathcal{S},\mathcal{T}$) of $\mathfrak {u}(N)_k$ Wess-Zumino conformal field theory.
The explicit form of Hopf link invariant on $S^{3}/\mathbb{Z}_{p}$ with representations $\mathcal{R}_{1}$ and $\mathcal{R}_{2}$ on the two components is given by 
\begin{align}
\widetilde{{\mathscr{V}}}^{(2,2)}_{\mathcal{R}_1,\mathcal{R}_2}\left(S^{3}/\mathbb{Z}_{p},k\right)&=\sum_{\mathcal{R}} \mathcal{S}_{\mathcal{R}_1\mathcal{R}} \mathcal{S}_{\mathcal{R}_2 \mathcal{R}}\mathcal{T}^{-p}_{\mathcal{R}\mathcal{R}}=\sum_{\mathcal{R}}\frac{\mathcal{S}_{\mathcal{R}_1\mathcal{R}}}{\mathcal{S}_{0\mathcal{R}}}\frac{\mathcal{S}_{\mathcal{R}_2\mathcal{R}}}{\mathcal{S}_{0\mathcal{R}}}\mathcal{S}_{0\mathcal{R}}^2 \mathcal{T}^{-p}_{\mathcal{R}\mathcal{R}}.\label{HopfLinkINV}
\end{align}
The summation runs over all the irreducible integrable representations $\mathcal{R}$ of $\mathfrak{u}(N)_k$. Considering $\mathcal{R}_{1}$ and $\mathcal{R}_{2}$ to be trivial representations we obtain the partition function of Chern-Simons theory on $S^{3}/\mathbb{Z}_{p}$
\begin{align}
\mathcal{Z}(S^{3}/\mathbb{Z}_{p},k)&=\sum_{\mathcal{R}}\mathcal{S}_{0\mathcal{R}}^{2}\mathcal{T}_{\mathcal{R}\mathcal{R}}^{-p}.
\label{eq:partitionfunction}
\end{align}
In the double scaling limit (\ref{eq:dslimit1}) the partition function is dominated by a single representation. Contributions due to other representations are suppressed by $\mathcal{O}(1/N^2)$. This is the standard saddle point approximation. The Hopf link invariant defined in (\ref{HopfLinkINV}) is unnormalised. We can define the normalised invariant as
\begin{align}\label{eq:normHopf}
{{\mathscr{V}}}^{(2,2)}_{\mathcal{R}_1,\mathcal{R}_2}\left(S^{3}/\mathbb{Z}_{p},k\right) = \frac{\widetilde{{\mathscr{V}}}^{(2,2)}_{\mathcal{R}_1,\mathcal{R}_2}\left(S^{3}/\mathbb{Z}_{p},k\right)}{\mathcal{Z}(S^{3}/\mathbb{Z}_{p},k)}
\end{align}
such that ${{\mathscr{V}}}^{(2,2)}_{\mathcal{R}_1,\mathcal{R}_2}\left(S^{3}/\mathbb{Z}_{p},k\right)$ with the representations $\mathcal{R}_{1}$ and $\mathcal{R}_{2}$ being trivial becomes identically equal to one. Note that, with $p=1$, ${\mathscr{V}}^{(2,2)}_{\mathcal{R}_{1},\mathcal{R}_{2}}(S^3/\mathbb{Z}_{p},k)$ becomes a Hopf link (\ref{U(N)invAriNantTof2,2mToruSliNK}) embedded inside three manifold $S^3$ with framing of component knots $f=\{1,1\}$. 

In order to proceed further we explicitly write down the modular transformation matrix $\mathcal{S}$ for $ {\mathfrak u}(N)_k$
\begin{align}\label{eq:Sab}
\mathcal{S}_{\mathcal{R}_{a}\mathcal{R}_{b}}&=\frac{(-i)^{N(N-1)/2}}{(k+N)^{N/2}}e^{\frac{\textstyle 2\pi i n^{(a)} n^{(b)}}{\textstyle N(N+k)}}\det M(\mathcal{R}_{a},\mathcal{R}_{b}),
\end{align}
where for any integrable representations $\mathcal{R}_a$ and $\mathcal{R}_b$ 
\begin{align}
    M_{ij}(\mathcal{R}_{a},\mathcal{R}_{b}) &=\exp\bigg[\frac{2\pi i}{k+N}\phi_{i}(\mathcal{R}_{a})\phi_{j}(\mathcal{R}_{b})\bigg],\quad i,j=1,..., N \\
 \quad \phi_{i}(\mathcal{R}_{a}) & = n_{i}^{(a)}-\frac{n^{(a)}}{N}-i+\frac{N+1}{2}.
\end{align}
Here, as before, $n^{(a/b)}$ and $n^{(a/b)}_i$ denote the total number of boxes and the number of boxes in the $i^{\text{th}}$ row corresponding to the Young diagram of $\mathcal{R}_{a/b}$ respectively.

The other modular transformation matrix $\mathcal{T}$ is given by
\begin{equation}
\mathcal{T}_{\mathcal{R}_{a}\mathcal{R}_{b}} =\exp \bigg[{2\pi i\big(\mathcal{Q}_{\mathcal{R}_{a}}-\frac{c}{24}\big)}\bigg]\delta_{\mathcal{R}_{a}\mathcal{R}_{b}},
\end{equation}
$\mathcal{Q}_{\mathcal{R}}$ and $c$ denote the conformal weight of representation $\mathcal{R}$ and central charge respectively
\begin{equation}
    \mathcal{Q}_{\mathcal{R}}=\frac{\kappa_{\mathcal{R}}}{(k+N)} \quad \text{and} \quad c=\frac{N(Nk+1)}{N+k}. 
\end{equation}

Next we want to calculate the ratio of modular transformation matrices $\frac{\mathcal{S}_{\mathcal{R}\mathcal{R}_{a}}}{\mathcal{S}_{0\mathcal{R}}}$ appearing in equation (\ref{HopfLinkINV}). Instead of representing a Young diagram by the set of box numbers $\{n_i^{(a)}\}$ we use the variables $\theta_i$ defined in equation (\ref{eq:theta-a}). 
Since in the double scaling limit $\theta_i$ ranges between $-\pi$ and $\pi$, we can arrange them as eigenvalues of an $N\times N$ unitary matrix 
\begin{equation}
    U = \text{diag}(e^{i\theta_1}, \cdots, e^{i\theta_N}).
\end{equation}
After some calculation one can show that the ratio $\frac{\mathcal{S}_{\mathcal{R}\mathcal{R}_{a}}}{\mathcal{S}_{0\mathcal{R}}}$ can be expressed as 
\begin{align}
\frac{\mathcal{S}_{\mathcal{R}\mathcal{R}_{a}}}{\mathcal{S}_{0\mathcal{R}}} = \mathbf{s}_{\mathcal{R}_{a}}(U), \label{eq:ratioModtransMatrices}
\end{align}
where the Schur polynomial $\mathbf{s}_{\mathcal{R}_{a}}(U)$ of the unitary group element $U$ in representation $\mathcal{R}_a$ is
\begin{align}
    \mathbf{s}_{\mathcal{R}_{a}}(U)&=\frac{\textstyle\det \left[e^{i\theta_{i}h_{j}^{(a)}}\right]}{\det\left[e^{i\theta_{i}(N-j)}\right]}.
\end{align}
 Note that the notation $U$ used  in equation (\ref{eq:ratioModtransMatrices}) corresponds to representation $\mathcal{R}$ on its left hand side i.e. the diagonal elements of $U$ correspond to $h_i$ (\ref{eq:h-a}) of $\mathcal{R}$. The unnormalised Hopf link invariant, therefore can be expressed as\footnote{Note the difference between the expressions of Hopf link here and in \cite{chakraborty2022large}.}
\begin{equation}\label{eq:Whopf}
\widetilde{{\mathscr{V}}}^{(2,2)}_{\mathcal{R}_{1},\mathcal{R}_{2}}(S^{3}/\mathbb{Z}_{p},k)=q^{\frac{pN(Nk+1)}{24}} \sum_{\mathcal{R}} q^{-{p} \kappa_{\mathcal{R}} } ~\mathcal{S}_{0\mathcal{R}}^{2} \  \mathbf{s}_{\mathcal{R}_{1}}(U) \mathbf{s}_{\mathcal{R}_{2}}(U),
\end{equation}
where $q$ is the $(k+N)$th root of unity (\ref{VariableS}). In equation (\ref{eq:Whopf}) each $\mathcal{R}$ corresponds to a unitary matrix $U$ in an ensemble of unitary matrices. The summation over $\mathcal{R}$ therefore corresponds to summation over unitary matrices. The factor $\mathcal{S}_{0\mathcal{R}}^2$ when written in terms of $\theta_i$ variables, reduces to Vandermonde measure for unitary matrices. Thus in the large $N$ limit we can write
\begin{equation}\label{eq:RtoU}
    \sum_{\mathcal{R}} \mathcal{S}_{0\mathcal{R}}^2 = \int \prod_i d\theta_i \prod_{i<j} \sin^2\left(\frac{\theta_i - \theta_j}{2}\right) \equiv \int [DU]
\end{equation}
where we have ignored few $\theta$ independent overall factors which doesn't affect our analysis. Hence the partition function (\ref{eq:partitionfunction}) can be written as a $0$-dimensional unitary matrix model with weight factor $e^{-(N+k)^2 S[\theta]}$ as
\begin{align}
    \label{eq:pfU}
   \mathcal{Z}(S^{3}/\mathbb{Z}_{p},k) = \int \prod_i d\theta_i \prod_{i<j} \sin^2\left(\frac{\theta_i - \theta_j}{2}\right)e^{-(N+k)^2 S[\theta]},
\end{align}
where
\begin{equation}\label{eq:weight}
    S[\theta] = \frac{p\lambda}{N\pi} \sum_{i=1}^N \left(\frac{\theta_i^2}{4} - \frac{\pi^2}{12}
    \right) + \frac{\pi p \lambda(1-\lambda)}{12}.
\end{equation}
Note that, in order to arrive at the above result we have replaced $p$ in equation (\ref{eq:partitionfunction}) by $-ip$. 
The normalised Hopf link invariant (\ref{eq:normHopf}) is therefore given as the integral of two Schur polynomials $ \mathbf{s}_{\mathcal{R}_1}(U)$ and $ \mathbf{s}_{\mathcal{R}_2}(U)$ over the unitary ensemble with the aforementioned weight factor
\begin{align}
{{\mathscr{V}}}^{(2,2)}_{\mathcal{R}_{1},\mathcal{R}_{2}}(S^{3}/\mathbb{Z}_{p},k) 
    & = \frac{1}{\mathcal{Z}(S^{3}/\mathbb{Z}_{p},k)} \int [DU] e^{-(N+k)^2 S[\theta]}  \mathbf{s}_{\mathcal{R}_1}(U) \mathbf{s}_{\mathcal{R}_2}(U)\nonumber \\
    & = \left\langle   \mathbf{s}_{\mathcal{R}_1}(U) \mathbf{s}_{\mathcal{R}_2}(U) \right\rangle.
    \label{eq:Hopfcorrel}
\end{align}
For large $\mathcal{R}_{1}$ and $\mathcal{R}_{2}$ (i.e. when the number of boxes $n_i^{(1)}$ and $n_{i}^{(2)}$ become of order $N$ or $k$, in the double scaling limit) it is difficult to calculate the correlation function even using the saddle point approximation. However, if $\mathcal{R}_{1}$ and $\mathcal{R}_{2}$ are small, one can find the leading order contribution to ${\mathscr{V}}^{(2,2)}_{\mathcal{R}_{1},\mathcal{R}_{2}}(S^{3}/\mathbb{Z}_{p},k)$ by computing $\mathbf{s}_{\mathcal{R}_{1}}(U)$ and $\mathbf{s}_{\mathcal{R}_{2}}(U)$ on the classical solution that extremizes the partition function (\ref{eq:partitionfunction}).

\subsection{Leading contribution to \texorpdfstring{$(L,L)$}{Lg} torus link invariant for small representations}\label{sec:largesymmrep}

For a (2,2) torus link invariant when the ``size'' of $\mathcal{R}_{1}$ and $\mathcal{R}_{2}$ placed on the component knots are much smaller than $N$, they do not back react on the classical solution extremizing the partition function. Hence we can consider the operators $\mathbf{s}_{\mathcal{R}_{1}}(U)$ and $~\mathbf{s}_{\mathcal{R}_{2}}(U)$ as probe and evaluate them on the background solution (\ref{eq:bgsol}).

Let us define a distribution function for the variables $\{\theta_i\}$ in the large $N$ limit as
\begin{equation}
    \rho(\theta) = \frac{1}{N}\sum_{i=1}^N \delta(\theta - \theta_i), \quad \text{such that} \quad \int_{-\pi}^\pi d\theta \rho(\theta) =1.\label{eq:evdensity}
\end{equation}
It was shown in Ref. \cite{Chakraborty:2021oyq} that in the double scaling limit (\ref{eq:dslimit1}) the partition function (\ref{eq:partitionfunction}) is dominated by the following distribution function\footnote{In this paper we have used the no-cap solution. In the double scaling limit CS theory admits a phase transition at some critical value of $\lambda$. Beyond that critical value the dominant distribution is given by a capped eigenvalue density. As a results the behaviour of correlation functions and hence knot/link invariants will change. We have commented on this in the conclusion section.}
\begin{align}
    \label{eq:bgsol}
    \rho(\theta)&=\frac{p}{2\pi^2\lambda}\tanh^{-1}\sqrt{1-\frac{e^{-2\pi \lambda/p}}{\cos^{2}(\theta/2)}},\quad \text{with}~-2\sec^{-1}e^{\pi \lambda/p}<\theta < 2 \sec^{-1}e^{\pi \lambda/p}
\end{align}
which is a solution to the following saddle point equation obtained from the partition function (\ref{eq:pfU})
    \begin{equation}\label{eq:sad0}
          \dashint \rho(\theta')\cot\left(\frac{\theta-\theta'}{2}\right)d\theta' = \frac{p}{2\pi\lambda}\theta.
\end{equation}
Therefore in the probe approximation the Hopf link invariant (\ref{eq:Hopfcorrel}) can be computed by evaluating the Schur polynomials on the solution (\ref{eq:bgsol})
\begin{equation}
{\mathscr{V}}^{(2,2)}_{\mathcal{R}_{1},\mathcal{R}_{2}}(S^{3}/\mathbb{Z}_{p},\lambda) 
    = \mathbf{s}_{\mathcal{R}_1}(\bar U) \mathbf{s}_{\mathcal{R}_2}(\bar U)\label{LHopfSMRep}
\end{equation}
where $\bar U$ corresponds to the distribution (\ref{eq:bgsol}).

The Schur polynomial admits a character expansion \cite{marino2005chern}
\begin{align}\label{eq:chaexp}
\mathbf{s}_{\mathcal{R}_a}(U) = \sum_{\Vec{k}^{(a)}} \frac{\chi_{\mathcal{R}_a}(c(\Vec{k}^{(a)}))}{z_{\Vec{k}^{(a)}}} \Upsilon_{\Vec{k}^{(a)}}(U),
\end{align}
where
\begin{align}
   \Vec{k}^{(a)}&=\left( k^{(a)}_{1},k^{(a)}_{2},k^{(a)}_3,...\right)~\text{such that}~\sum_{r}rk^{(a)}_{r}=n^{(a)}, \label{eq:chacond}
\end{align}
and
\begin{align}
\Upsilon_{\Vec{k}^{(a)}}(U) = \prod_{r} \left(\text{Tr}~ U^r\right)^{k^{(a)}_r}, \quad z_{\Vec{k}^{(a)}} = \prod_r k^{(a)}_r !~ r^{k^{(a)}_r}.
\label{eq:upsilon}
\end{align}
$n^{(a)}$ being the total number of boxes in representation $\mathcal{R}_{a}$. $c(\Vec{k}^{(a)})\equiv 1^{k^{(a)}_1} 2^{k^{(a)}_2}3^{k^{(a)}_{3}} \cdots$ is the conjugacy class of the symmetric group $\mathbb{G}_{{n^{(a)}}}$ of degree $n^{(a)}$, consisting of $k^{(a)}_1$ cycles of length 1, $k^{(a)}_2$ cycles of length 2 and so on. In other words, $k^{(a)}_{r}$ denotes the number of $r$-cycle of the conjugacy class $c(\Vec{k}^{(a)})$.  
$\chi_{\mathcal{R}_a}(c(\Vec{k}^{(a)}))$ is the corresponding character, in representation $\mathcal{R}_a$. 

Since to each vector $\Vec{k}^{(a)}$ we can associate a conjugacy class $c(\Vec{k}^{(a)})$, the sum over $\Vec{k}^{(a)}$ in (\ref{eq:chaexp}) represents a sum over all possible conjugacy classes subjected to the condition (\ref{eq:chacond}). Thus in the probe approximation, computation of the Hopf link invariant reduces to computation of the quantities $\Upsilon_{\Vec{k}^{(a)}}(U)$ on the solution (\ref{eq:bgsol}).

In the large $N$ limit the expectation value of $\Tr U^r$ is dominated by the background solution i.e
\begin{equation}
    \langle \Tr U^r \rangle \rightarrow {\Tr}\bar U^r = N \int e^{i r \theta}  \rho(\theta) d\theta = N \int \cos r\theta ~\rho(\theta) d\theta \equiv N \rho_r,
\end{equation}
where 
\begin{align}
    \rho_{r}&=\int\cos r\theta ~\rho(\theta) d\theta,\label{eq:rhordef}
\end{align}
with $\rho(\theta)$ as given in (\ref{eq:bgsol}).  Therefore, in the large $N$ limit $\mathbf{s}_{\mathcal{R}_{a}}(\bar U)$ is given by

\begin{equation}
    \begin{aligned}
        \label{eq:Schur}
\mathbf{s}_{\mathcal{R}_{a}}(\bar U) &= \sum_{\Vec{k}^{(a)}} \frac{\chi_{\mathcal{R}_a}(c(\Vec{k}^{(a)}))}{z_{\Vec{k}^{(a)}}} \prod_{r} N^{k^{(a)}_r} \rho_r^{k^{(a)}_r} = \sum_{\Vec{k}^{(a)}} N^{K^{(a)}}  \frac{\chi_{\mathcal{R}_a}(c(\Vec{k}^{(a)}))}{z_{\Vec{k}^{(a)}}} \prod_r \rho_{r}^{k^{(a)}_r}, \\
~&~~~~~~~~~~\text{where}~\quad K^{(a)} = \sum_r k^{(a)}_r.
    \end{aligned}
\end{equation}
For small representations $k^{(a)}_r$'s are $\mathcal{O}(1)$ numbers. Hence $z_{\Vec{k}^{(a)}}$'s are also $\mathcal{O}(1)$ numbers. As a result the $N$ dependence of $\mathbf{s}_{\mathcal{R}_{a}}(\bar U)$ is given by $N^{K^{(a)}}$. Therefore, in the large $N$ limit $\mathbf{s}_{\mathcal{R}_{a}}(\bar U)$ is dominated by a conjugacy class for which $K^{(a)}$ is maximum subjected to (\ref{eq:chacond}). Using (\ref{eq:chacond}) we can replace $k^{(a)}_1$ in $K^{(a)}$ and find
\begin{equation}
K^{(a)} = n^{(a)} - k^{(a)}_2 -2k^{(a)}_3 -\cdots.    
\end{equation}
Hence, the maximum value of $K^{(a)}$ is given by $k^{(a)}_1 = n^{(a)}$ and  $k^{(a)}_2=k^{(a)}_3=\cdots =0$. This choice reduces equation (\ref{eq:chacond}) to $\Vec{k}^{(a)}=\left(n^{(a)},0,0,... \right)$, as a result the Schur polynomial (\ref{eq:Schur}) evaluated on the classical solution becomes
\begin{align}
\mathbf{s}_{\mathcal{R}_{a}}(\bar U) &=N^{n^{(a)}}\rho_{1}^{n^{(a)}}\frac{\chi_{\mathcal{R}_{a}}(1^{n^{(a)}})}{(n^{(a)})!},\label{SchUrSMrep}
\end{align} 
where $\rho_{1}$ can be evaluated using (\ref{eq:rhordef}) and is given by
\begin{equation}\label{eq:TrU0}
    \rho_1 = p\left(\frac{1-e^{-\frac{2\pi \lambda}{p}}}{2\pi\lambda}\right),
\end{equation}
$\chi_{\mathcal{R}_a}\left(1^{n^{(a)}}\right)$ is the character of the conjugacy class consisting of $n^{(a)}$ number of 1-cycles in representation $\mathcal{R}_a$. Thus the leading contribution (denoted by $\boldsymbol{\ell}.\boldsymbol{c}.$) to the unknot invariant in fundamental representation is given by
\begin{equation}
\boldsymbol{\ell}.\boldsymbol{c}.[{\mathscr{V}}_{\Box}^{\left(2,1\right)}\left( S^{3}/\mathbb{Z}_{p},\lambda\right)]=\mathbf{s}_{\Box}(\bar{U}) = N \rho_1,
\end{equation}
The superscript $(2,1)$ refers to an unknot obtained as closure of a two strand braid with one crossing. Therefore, the leading contribution to the Hopf link invariant  in the double scaling limit can be expressed as (\ref{LHopfSMRep})
\begin{align}
\boldsymbol{\ell}.\boldsymbol{c}.\left[{\mathscr{V}}^{(2,2)}_{\mathcal{R}_{1},\mathcal{R}_{2}}(S^{3}/\mathbb{Z}_{p},\lambda) \right]
   & = N^{n^{(1)} +n^{(2)}} \rho_1^{n^{(1)}+n^{(2)}} \frac{\chi_{\mathcal{R}_1}\left(1^{n^{(1)}}\right)\chi_{\mathcal{R}_2}\left(1^{n^{(2)}}\right)}{\left( n^{(1)}\right)! \left(n^{(2)}\right)!}\nonumber\\
     = \bigg( \boldsymbol{\ell}.\boldsymbol{c}.&\left[{\mathscr{V}}_{\Box}^{\left(2,1\right)} \left( S^{3}/\mathbb{Z}_{p},\lambda\bigg)\right]\right)^{n^{(1)} + n^{(2)}} \frac{\chi_{\mathcal{R}_1}\left(1^{n^{(1)}}\right)\chi_{\mathcal{R}_2}\left(1^{n^{(2)}}\right)}{\left( n^{(1)}\right)! \left(n^{(2)}\right)!}\label{LeadingContofHopf}
\end{align}
It shows that the leading contribution of the Hopf link invariant with small representation placed on the component knots is proportional to appropriate powers of the leading contribution of unknot invariant in fundamental representation. Likewise, the leading contribution of unknot carrying arbitrary representation is
\begin{align}
\boldsymbol{\ell}.\boldsymbol{c}.  \left[ {\mathscr{V}}_{\mathcal{R}_1}^{\left(2,1\right)} (S^{3}/\mathbb{Z}_{p},\lambda) \right]& = \left( \boldsymbol{\ell}.\boldsymbol{c}.\left[{\mathscr{V}}_{\Box}^{\left(2,1\right)}(S^3/\mathbb{Z}_{p},\lambda) \right]\right)^{n^{(1)}} \frac{\chi_{\mathcal{R}_1}\left(1^{n^{(1)}}\right)}{\left( n^{(1)}\right)!}.\label{LeadingContofUnknot}
\end{align}
We can generalise (\ref{LeadingContofHopf}) for $(L,L)$ torus link invariant in the probe approximation and it is given by
\begin{align}
\boldsymbol{\ell}.\boldsymbol{c}.\left[{\mathscr{V}}^{(L,L)}_{\mathcal{R}_{1},\cdots \mathcal{R}_{L}}(S^{3}/\mathbb{Z}_{p},\lambda)\right] = \prod_{a=1}^L N^{n^{(a)}} \rho_1^{n^{(a)}} ~\frac{\chi_{\mathcal{R}_a}\left(1^{n^{(a)}}\right)}{\left(n^{(a)} \right)!}\equiv \prod_{a=1}^{L}\boldsymbol{\ell}.\boldsymbol{c}. \left[ {\mathscr{V}}_{\mathcal{R}_a}^{\left(2,1\right)} (S^{3}/\mathbb{Z}_{p},\lambda) \right].\label{l.c(L,L)}
\end{align}
For $(3,3)$ torus link, with the three component knots carrying  some specific representations with small number of boxes in the corresponding Young diagrams, we tabulate the colored HOMFLY-PT polynomials and the leading contributions in Table \ref{Tableof3,3lc}. Note that in order to match the matrix model result (\ref{l.c(L,L)}) with the tabulated ones we need to follow the \textit{prescription} as mentioned below:
\begin{itemize}
    \item[$\bullet$] first, replace $p \rightarrow i p$ in the expressions obtained by matrix model calculations to compensate the replacement $p \rightarrow -i p$ done earlier (see below equation (\ref{eq:weight})),
    \item[$\bullet$] then set $p=1$, since the colored HOMFLY-PT polynomials are obtained for framed links embedded inside $S^3$.
\end{itemize} 
Explicitly, it implies that 
\begin{align}
\mathscr{V}^{(L,L)}_{\mathcal{R}_{1},...,\mathcal{R}_{L}}(S^3/\mathbb{Z}_{p},\lambda)\overset{\substack{p\rightarrow ip,\\p=1}}{\longrightarrow}\mathcal{W}^{[(L,L),f=\{1,...,1\}]}_{\mathcal{R}_{1},...,\mathcal{R}_{L}}(S^3,\lambda).\label{eq:MaPPing}
\end{align}
When these modifications are taken into account, the expression of $\rho_{1}$ (\ref{eq:TrU0}) becomes identical to the leading order contribution of the vertical framing $U(N)$ invariant $\mathcal{W}_{\Box}^{(2,1)}(S^3,\lambda)$ of an unknot with fundamental representation placed on its component. 

The leading contributions as obtained from the colored HOMFLY-PT polynomials of $(3,3)$ torus links (refer to Table \ref{Tableof3,3lc}) are in exact agreement with those (\ref{l.c(L,L)}) obtained using the eigenvalue density ($\ref{eq:bgsol}$) extremising the partition function. It is to be noted that, in the large $N$ limit the evaluation of leading contribution using the eigenvalue density of the dominant representation is only possible when the representations placed on the component knots of the link are much smaller as compared to the rank $N$ of the gauge group at consideration.
\\
\\

{\footnotesize
\begin{longtable}{ |p{2.7cm}|p{9cm}|p{2.1cm}| }
 \hline
\begin{center}
$\mathcal{W}_{\mathcal{R}_{1},\mathcal{R}_{2},\mathcal{R}_{3}}^{(3,3)}(q,v)$
\end{center} & \begin{center}
    The colored HOMFLY-PT polynomial
\end{center}& \begin{center}
    Leading contribution \end{center}\\
 \hline
 \begin{center}
   $\mathcal{W}^{(3,3)}_{\ydiagram[]{1},\ydiagram[]{1},\ydiagram[]{1}}$
 \end{center} &   $(-1+q)^{-3}q^{-3/2}\big(-1 + 2 q - 2 q^2 + q^3 - 2 q^4 + 2 q^5 - q^6 + v - q v + q^2 v +$
 $ 
 q^3 v + q^4 v 
     - q^5 v + q^6 v - q v^2 + q^2 v^2 - 3 q^3 v^2 + 
 q^4 v^2 - q^5 v^2 + q^3 v^3\big)$  & \begin{center}
     $\frac{i(-1+e^{2\pi i \lambda})^{3}N^3}{8\pi^3 \lambda^3}$
 \end{center}\\
 \hline
\begin{center}
    $\mathcal{W}_{\ydiagram[]{2},\ydiagram[]{1},\ydiagram[]{1}}^{(3,3)}$
\end{center} & $(-1 + q)^{-4}q^{-2}(1+q)^{-1} \big(1 - 2 q + q^2 + q^3 - q^4 + q^5 - q^7 + q^8 - v + q v - q^3 v - 
 q^4 v - q^6 v - q^9 v + q v^2 - q^2 v^2 + q^3 v^2 +$ $ q^4 v^2 + 
 q^5 v^2 + q^6 v^2 + q^7 v^2 + q^9 v^2 - q^3 v^3 + q^4 v^3 - 
 2 q^5 v^3 - q^6 v^3 - q^8 v^3 + q^6 v^4\big)$ & \begin{center}
      $\frac{(-1+e^{2\pi i \lambda})^{4}N^4}{32 \pi^{4}\lambda^4}$
  \end{center}\\
 \hline
  \begin{center}
      $\mathcal{W}_{\ydiagram[]{2},\ydiagram[]{2},\ydiagram[]{1}}^{(3,3)}$
  \end{center}& $(-1+q)^{-5}q^{-7/2}(1+q)^{-2} \big( -1 + 2 q - 3 q^3 + 2 q^4 + q^5 - 3 q^6 + 2 q^7 + q^8 - 
 3 q^9 + q^{10} + q^{11} - q^{12} + v - q v - q^2 v + 2 q^3 v + q^6 v - 
 q^7 v + q^8 v + 2 q^9 v - q^{11} v + q^{12} v + q^{13} v - q v^2$ $ + 
 q^2 v^2 - q^4 v^2 - q^5 v^2 - q^7 v^2 - q^8 v^2 - 2 q^9 v^2 - 
 2 q^{10} v^2 + q^{11} v^2 - q^{12} v^2 - 2 q^{13} v^2 + q^3 v^3 - q^4 v^3 + 
 3 q^6 v^3 - q^7 v^3 + 5 q^9 v^3 + 2 q^{12} v^3 + q^{13} v^3 - q^6 v^4 + 
 q^7 v^4 - q^8 v^4 - 2 q^9 v^4 - q^{11} v^4 - q^{12} v^4 + q^{10} v^5\big)$ &\begin{center}
     $-\frac{i(-1+e^{2\pi i \lambda})^{5}N^5}{128 \pi^5 \lambda^5}$
 \end{center}\\
 \hline
\begin{center}
      $\mathcal{W}_{\ydiagram[]{2},\ydiagram[]{2},\ydiagram[]{2}}^{(3,3)}$
\end{center}& $(-1+q)^6 q^{-6}(1+q)^{-3}\big(1 - 2 q - q^2 + 5 q^3 - 2 q^4 - 4 q^5 + 4 q^6 + q^7 - 3 q^8 + 
 2 q^{10} + q^{12} - 2 q^{13} - 2 q^{14} + 4 q^{15} - 2 q^{17} + q^{18} - v + q v + 
 2 q^2 v - 3 q^3 v - q^4 v + 3 q^5 v - 2 q^6 v - q^7 v + 3 q^8 v - 
 2 q^9 v - 2 q^{10} v + q^{11} v - 3 q^{12} v + q^{13} v + 4 q^{14} v - 
 4 q^{15} v - 3 q^{16} v + 2 q^{17} v - q^{19} v + q v^2 - q^2 v^2 - $ $
 q^3 v^2 + 2 q^4 v^2 + q^7 v^2 - q^8 v^2 + q^9 v^2 + 2 q^{10} v^2 + 
 2 q^{11} v^2 + q^{12} v^2 + 2 q^{13} v^2 - q^{14} v^2 + 5 q^{16} v^2 + 
 q^{17} v^2 - q^{18} v^2 + 2 q^{19} v^2 - q^3 v^3 + q^4 v^3 + q^5 v^3 - 
 3 q^6 v^3 + 2 q^8 v^3 - 3 q^9 v^3 - q^{10} v^3 - 5 q^{12} v^3 - $ $
 2 q^{13} v^3 + q^{14} v^3 - 3 q^{15} v^3 - 3 q^{16} v^3 - q^{17} v^3 - 
 2 q^{18} v^3 - q^{19} v^3 + q^6 v^4 - q^7 v^4 - q^8 v^4 + 4 q^9 v^4 - 
 4 q^{11} v^4 + 6 q^{12} v^4 + 4 q^{13 }v^4 - 2 q^{14} v^4 + 2 q^{15} v^4 + 3 q^{16} v^4 + q^{17} v^4 + 2 q^{18} v^4 - q^{10} v^5 + q^{11} v^5 - 
 3 q^{13} v^5 - 2 q^{16} v^5 - q^{17} v^5 + q^{15} v^6 \big)$& \begin{center}
     $-\frac{(-1+e^{2\pi i\lambda})^6 N^6}{512 \pi^6\lambda^6}$
 \end{center}\\
 \hline
 \begin{center}
     $\mathcal{W}_{\ydiagram[]{2,1},\ydiagram[]{1},\ydiagram[]{1}}^{(3,3)}$
 \end{center} & $(-1+q)^{-5}q^{-5/2}(1+q+q^2)^{-1}\big( -q + 2 q^2 - 2 q^3 + 2 q^4 - 4 q^5 + 5 q^6 - 4 q^7 + 2 q^8 - 2 q^9 + 
 2 q^{10} - q^{11} + v - q v + q^2 v - q^3 v + 3 q^4 v - q^5 v + q^6 v - 
 q^7 v + 3 q^8 v - q^9 v + q^{10} v - q^{11} v + q^{12} v - v^2 + q v^2 - 
 2 q^2 v^2 + 2 q^3 v^2 - 6 q^4 v^2 + 4 q^5 v^2 - 6 q^6 v^2 + $ $
 4 q^7 v^2 - 6 q^8 v^2 + 2 q^9 v^2 - 2 q^{10} v^2 + q^{11} v^2 - 
 q^{12} v^2 + q v^3 - q^2 v^3 + 2 q^3 v^3 + 4 q^5 v^3 - 2 q^6 v^3 + 
 4 q^7 v^3 + 2 q^9 v^3 - q^{10} v^3 + q^{11} v^3 - q^3 v^4 + q^4 v^4 - 
 3 q^5 v^4 + q^6 v^4 - 3 q^7 v^4 + q^8 v^4 - q^9 v^4 + q^6 v^5\big)$ & \begin{center}
     $-\frac{i(-1+e^{2\pi i \lambda})^5 N^5}{96\pi^5 \lambda^5}$
 \end{center}\\
 \hline
 \begin{center}
     $\mathcal{W}_{\ydiagram[]{2,1},\ydiagram[]{2},\ydiagram[]{1}}^{(3,3)}$
 \end{center}& $(-1+q)^6 q^{-4}(1+q)^{-1}(1+q+q^2)^{-1}\big( q - 2 q^2 + q^3 + q^5 - 2 q^7 + 2 q^8 - q^9 + q^{10} - q^{12} + 2 q^{13} - 2 q^{14} + q^{15} - v + q v - q^4 v - q^5 v - 2 q^{10} v - q^{11} v - q^{16} v + v^2 - q v^2 + q^2 v^2 - q^3 v^2 + 2 q^4 v^2 + q^5 v^2 + q^6 v^2 + q^7 v^2 + 2 q^9 v^2 + 2 q^{10} v^2 + 2 q^{11} v^2 + 2 q^{12} v^2 - q^{13} v^2 +$ $ 2 q^{14} v^2 - q^{15} v^2 + 2 q^{16} v^2 - q v^3 + q^2 v^3 - q^3 v^3 - 2 q^5 v^3 - q^6 v^3 - 2 q^7 v^3 - q^8 v^3 - 2 q^9 v^3 - 3 q^{10} v^3 - 3 q^{11} v^3 - q^{12} v^3 - 2 q^{13} v^3 - q^{15} v^3 - q^{16} v^3 + q^3 v^4 - q^4 v^4 + q^5 v^4 + q^6 v^4 + 2 q^7 v^4 + q^8 v^4 + q^{9} v^4 + 3 q^{10} v^4+ 3 q^{11} v^4 + 2 q^{13} v^4 + q^{15} v^4 - q^6 v^5 + q^7 v^5 - 2 q^8 v^5 - 2 q^{10} v^5 -q^{11} v^5 - q^{13} v^5 + q^{10} v^6\big)$ & \begin{center}
     $-\frac{(-1+e^{2\pi i \lambda})^{6}N^6}{384 \pi^6 \lambda^6}$
 \end{center}\\
 \hline
 
\caption{{$(3,3)$ Torus link invariants and their leading contributions}}\label{Tableof3,3lc}
\end{longtable}
}
For a large representation $\mathcal{R}_{a}$, the total number of boxes $n^{(a)}$ is of $\mathcal{O}(N)$ which implies some of the $k^{(a)}_r$'s (\ref{eq:chacond}) can also become of $\mathcal{O}(N)$. As a result the denominator $z_{\Vec{k}}$ in equation (\ref{eq:chaexp}) will no longer be of $\mathcal{O}(1)$. Hence the above saddle point analysis will not give correct answer in such scenarios. However, we have a way to find the leading contribution for a \textit{class} of large representations which will be addressed in the following subsection.
\subsection{Leading contribution to \texorpdfstring{$(L,L)$}{Lg} torus link invariant for large symmetric representations}\label{sec:largesymmrep}
In this section, we provide a large $N$ expression for $(L,L)$ link invariant when all the representations are large symmetric representations. The $U(N)$ invariant for $(L,L)$ torus link, embedded inside three manifold $S^3/\mathbb{Z}_{p}$, carrying  integrable representations $\mathcal{R}_{1},...,\mathcal{R}_{L}$ on the component knots is given by
\begin{equation}
{\mathscr{V}}^{(L,L)}_{\mathcal{R}_{1},\cdots, \mathcal{R}_{L}}(S^{3}/\mathbb{Z}_{p},k) =\frac{1}{\mathcal{Z}(S^{3}/\mathbb{Z}_{p},k)}  \sum_{\mathcal{R}} \mathcal{S}_{0\mathcal{R}}^2 \mathcal{T}_{\mathcal{R}\mathcal{R}}^{-p}  \prod_{a=1}^L \mathbf{s}_{\mathcal{R}_a}(U)\equiv\left\langle   \prod_{a=1}^L \mathbf{s}_{\mathcal{R}_a}(U)\right\rangle.\label{L,LInvS^3/Z_p}
\end{equation}
Using the character expansion for Schur polynomial (\ref{eq:chaexp}) and (\ref{eq:RtoU}), in the double scaling limit we can express the above invariant as
\begin{equation}
\mathscr{V}^{(L,L)}_{\mathcal{R}_{1},\cdots \mathcal{R}_{L}}(S^{3}/\mathbb{Z}_{p},\lambda) =\frac{1}{\mathcal{Z}(S^{3}/\mathbb{Z}_{p},\lambda)} \sum_{\Vec{k}^{(1)}, \cdots \Vec{k}^{(L)}} \prod_{a=1}^L \frac{\chi_{\mathcal{R}_a}(c(\Vec{k}^{(a)}))}{z_{\Vec{k}^{(a)}}} \int [D\theta]e^{-N^2 S^{\left(\Vec{k}^{(1)}, \cdots, \Vec{k}^{(L)}\right)}_{eff}[\rho(\theta)]}
\end{equation}
where
\begin{align}
    S^{(\Vec{k}^{(1)}, \cdots, \Vec{k}^{(L)})}_{eff}[\rho(\theta)] & = - \frac{1}{2}\int d\theta \rho(\theta) \dashint d\theta'\rho(\theta')\log\left[ 4\sin^{2}\left( \frac{\theta-\theta'}{2}\right) \right]  \nonumber\\
        & +\frac{p}{\pi \lambda}\int \rho(\theta)\left( \frac{\theta^{2}}{4} - \frac{\pi^{2}}{12} \right) d \theta + \frac{\pi p (1-\lambda)}{12 \lambda} \nonumber\\
        & - \frac{1}{N^2}\sum_r (k_r^{(1)} + \cdots + k_r^{(L)}) \ln \left[N \int d\theta \rho(\theta) e^{i r \theta}\right].
        \end{align}
For a given $\Vec{k}^{(1)}, \cdots, \Vec{k}^{(L)}$, $\int [D\theta]e^{-N^2 S^{(\Vec{k}^{(1)}, \cdots, \Vec{k}^{(L)})}_{eff}[\rho(\theta)]}$ is dominated by a classical configuration satisfying the saddle point equation
\begin{equation}\label{eq:sad}
          \dashint \rho(\theta')\cot\left(\frac{\theta-\theta'}{2}\right)d\theta' = \frac{p}{2\pi\lambda}\theta - \frac{i}{N^2} \sum_r \frac{r (k_{r}^{(1)} + \cdots + k_{r}^{(L)})}{\rho_r} e^{i r \theta}.
\end{equation}
This is a coupled equation because the moments of the distribution $\rho_r = \int d\theta \rho(\theta) e^{i r \theta}$ appear in the last term and hence difficult to solve. The last term is proportional to $1/N^2$. Apparently, it seems that this term will not be important to find $\rho(\theta)$ in the large $N$ limit. But if any of the $\mathcal{R}_a$ is large such that $n^{(a)}$ is of $\mathcal{O}(N^2)$, then following (\ref{eq:chacond}) some of the $k^{(a)}_r$ can also be of $\mathcal{O}(N^2)$. In such cases we can not neglect the last term to solve for $\rho(\theta)$.

However, in the case of all representations ($\mathcal{R}_a$) large but completely symmetric, there can be at most $k$ boxes in the first row and hence $n^{(a)} \leq  k$. As a result all the $k_r^{(a)}$s are less than (or equal to) $k$. Therefore in the large $N$ limit we can still neglect the last term. This suggest us that in order to compute $\mathscr{V}^{(L,L)}_{\mathcal{R}_{1},\cdots \mathcal{R}_{L}}(S^{3}/\mathbb{Z}_{p},\lambda)$ for large symmetric representations we can still use the probe approximation and calculate the Schur polynomials on the background solution (\ref{eq:bgsol}). Thus the $(L,L)$ link invariant (\ref{L,LInvS^3/Z_p}) becomes
\begin{equation}
\mathscr{V}^{(L,L)}_{\mathcal{R}_{1},\cdots \mathcal{R}_{L}}(S^{3}/\mathbb{Z}_{p},\lambda) = \prod_{a=1}^L \mathbf{s}_{\mathcal{R}_{a}}(\bar U). \label{eq:LSymbarU}
\end{equation}
where $\bar{U}$ corresponds to the density distribution solving (\ref{eq:sad0}). Note that, unlike (\ref{SchUrSMrep}) $\mathbf{s}_{\mathcal{R}_a}(\bar U)$ will no longer be dominated by the conjugacy class having $n^{(a)}$ number of 1-cycles.

Since the character of any conjugacy class of the permutation group for a completely symmetric representation is 1, the Schur polynomial (\ref{eq:Schur}) can be written as,
\begin{align}
    \mathbf{s}_{\mathcal{R}_{a}}(\bar U) & = \sum_{\Vec{k}^{(a)}} \prod_{r} \frac{(N \rho_r)^{k^{(a)}_r}}{k^{(a)}_r!~ r^{k^{(a)}_r}}\nonumber \\
    & = \sum_{\Vec{k}^{(a)}} \exp\left[\sum_r \left( k^{(a)}_r \ln \left(\frac{N \rho_r}{r}\right) - k^{(a)}_r \ln k^{(a)}_r + k^{(a)}_r \right)\right]
    \end{align}
subjected to the condition ${\sum_{r}} r k^{(a)}_r = n^{(a)}$. We use the Lagrange multiplier $\alpha^{(a)}$ to extremise the above summation. Suppose
\begin{equation}
    \mathcal{H}^{(a)} = \sum_r \left(  k^{(a)}_{r} \ln \left(\frac{N \rho_r}{r}\right) - k^{(a)}_{r} \ln k^{(a)}_{r} + k^{(a)}_{r} \right)+ \alpha^{(a)}\left( \sum_r r k^{(a)}_r - n^{(a)} \right).
\end{equation}
Extremising $\mathcal{H}^{(a)}$ with respect to $k^{(a)}_r$ we find,
\begin{equation}
    k^{(a)}_r = \frac{N\rho_r}{r} e^{r \alpha^{(a)}}.
\end{equation}
The Lagrange multiplier $\alpha^{(a)}$ can be obtained by substituting $k^{(a)}_r$ in the equation $\sum_r r k^{(a)}_r = n^{(a)}$
\begin{equation}\label{eq:lagmul}
    \sum_r \rho_r e^{r \alpha^{(a)}} = \frac{n^{(a)}}{N}.
\end{equation}
Finally evaluating $\mathcal{H}^{(a)}$ on the solution we find
\begin{equation}
    \overline{\mathcal{H}}^{(a)} = \sum_r \frac{N\rho_r}{r}e^{r \alpha^{(a)}} - \alpha^{(a)} ~n^{(a)}.
\end{equation}
Hence we have
\begin{equation}
\mathbf{s}_{\mathcal{R}_{a}}(\bar U) = e^{\overline{\mathcal{H}}^{(a)}}.\label{eq:SchurSymL}
\end{equation}
We can use this result to find the Schur polynomial evaluated on the classical solution. Analytically it is challenging to find the Schur polynomial for any arbitrary values of $\lambda$. However, we can find it for small $\lambda$.

For small $\lambda$ the eigenvalue distribution (\ref{eq:bgsol}) is given by
\begin{equation}
    \rho(\theta) = \frac{p}{2\pi^2 \lambda}\sqrt{\frac{2\pi \lambda}{p} - \frac{\theta^2}{4}}.\label{eq:rhosmall}
\end{equation}
We can compute $\rho_r$ by using the formula (\ref{eq:rhordef}) with $\rho(\theta)$ as defined above and range of $\theta$ lying from $-2\sqrt{{2\pi \lambda}/{p}}$ to $2 \sqrt{{2\pi \lambda}/{p}}$. The result is as follows
\begin{equation}
    \rho_r = 1 - \frac{r^2 \pi \lambda}{p}.
\end{equation}
Using (\ref{eq:lagmul}) we can find that the Lagrange multiplier is given by,
\begin{eqnarray}
    \alpha^{(a)} = \ln\left(\frac{\kappa^{(a)}}{1+\kappa^{(a)}}\right) + \frac{(1+2\kappa^{(a)})\pi \lambda}{p}
\end{eqnarray}
where $\kappa^{(a)} = \frac{n^{(a)}}{N}$. Hence $\overline{\mathcal{H}}^{(a)}$ can be evaluated to obtain,
\begin{equation}
    \overline{\mathcal{H}}^{(a)} = \ln \mathbf{s}_{\mathcal{R}_a} = (1+\kappa^{(a)})\ln(1+\kappa^{(a)}) - \kappa^{(a)} \ln\kappa^{(a)} -\frac{\kappa^{(a)}(1+\kappa^{(a)})\pi \lambda}{p}.\label{eq:barH}
\end{equation}
Substituting $\overline{\mathcal{H}}^{(a)}$  in equation (\ref{eq:SchurSymL}) and then using (\ref{eq:LSymbarU})  we can deduce the leading order contribution for $(L,L)$ torus link invariant whose components carry arbitrary large symmetric representations.

To understand how different components are intertwined in a link we have to go beyond the leading contribution. From the matrix model point of view, the subleading correction is captured by the connected piece of the correlator. The computation of the connected piece becomes tedious involving resolvent and contour integration for a general $(L,L)$ torus link. However, one can extract the subleading contributions from the colored HOMFLY-PT polynomials (as in Table \ref{Tableof3,3lc}) when small representations are placed on the component knots of the torus link.

In order to confirm that the subleading contribution from the colored HOMFLY-PT matches with the matrix model approach,
we will explicitly present the  computation of connected correlator for the Hopf link invariant in the following section.
\section{Subleading corrections to Hopf link invariant}
\label{sec:subleading}
In this section we compute the subleading contribution to the invariant of Hopf link. From (\ref{eq:Hopfcorrel}) we note that the Hopf link invariant is equal to correlation of two Schur polynomials. Let us consider that both the component knots of the link are in fundamental representation i.e. ${\cal R}_1 = {\cal R}_2 = \Box$. The character expansion of the Schur polynomial (\ref{eq:chaexp}) tells us that $\mathbf{s}_{\Box}(U) = \Tr U$. Thus the Hopf link invariant with both the components in fundamental representations can be expressed as
\begin{eqnarray}
\mathscr{V}^{(2,2)}_{\Box,\Box}(S^{3}/\mathbb{Z}_{p},k) = \langle \Tr U \Tr U\rangle .
\end{eqnarray}
Likewise, an unknot invariant in fundamental representation is given by 
\begin{align}
    \mathscr{V}^{(2,1)}_{\Box}(S^{3}/\mathbb{Z}_{p},k) = \langle \Tr U\rangle. 
\end{align}
We introduce a rescaled trace operator
\begin{equation}
    \mathscr{T}\mathrm{r} U^m = \frac{1}{N} \Tr U^m, \quad m \in \mathbb{Z} 
\end{equation}
such that $ \langle \mathscr{T}\mathrm{r} U^m \rangle$, when evaluated on saddle point gives a $\mathcal{O}(1)$ number, i.e. the leading contribution to $ \langle \mathscr{T}\mathrm{r} U^m \rangle$ becomes of $\mathcal{O}(1)$. The corrections to $ \langle \mathscr{T}\mathrm{r} U^m \rangle$ are suppressed by higher powers of $N$. In terms of rescaled trace operators, the Hopf link and unknot invariants can be written as
\begin{equation}\label{eq:HUrescale}
\mathscr{V}^{(2,2)}_{\Box,\Box}(S^{3}/\mathbb{Z}_{p},k) = N^2 \langle  \mathscr{T}\mathrm{r} U  \mathscr{T}\mathrm{r} U\rangle, \quad \mathscr{V}^{(2,1)}_{\Box}(S^{3}/\mathbb{Z}_{p},k) = N \langle  \mathscr{T}\mathrm{r} U\rangle.
\end{equation} 

We define the \emph{connected piece} of a two-point function in the following way \cite{eynard2015random},
\begin{equation}\label{eq:connectdef}
    \langle  \mathscr{T}\mathrm{r} U  \mathscr{T}\mathrm{r} U\rangle_{c} \equiv N^2 \left(\langle  \mathscr{T}\mathrm{r} U  \mathscr{T}\mathrm{r} U\rangle - \langle  \mathscr{T}\mathrm{r} U \rangle^2 \right),
\end{equation}
where we have put an overall factor of $N^2$ on the right hand side such that the leading contribution of $\langle  \mathscr{T}\mathrm{r} U  \mathscr{T}\mathrm{r} U\rangle_{c}$ is again of $\mathcal{O}(1)$. This connected piece  can be expressed in terms of the Hopf link and unknot invariants as
\begin{equation}\label{eq:hopf-unsq}
    \mathscr{V}^{(2,2)}_{\Box,\Box}(S^{3}/\mathbb{Z}_{p},k) - \left( \mathscr{V}^{(2,1)}_{\Box}(S^{3}/\mathbb{Z}_{p},k) \right)^2 = \langle  \mathscr{T}\mathrm{r} U  \mathscr{T}\mathrm{r} U\rangle_{c}.
\end{equation}
In the large $N$ limit $\langle  \mathscr{T}\mathrm{r} U\rangle $ admits the expansion
\begin{equation}
    \langle  \mathscr{T}\mathrm{r} U\rangle = \langle  \mathscr{T}\mathrm{r} U \rangle_0 + \frac{1}{N^2} \langle  \mathscr{T}\mathrm{r} U \rangle_1 + \mathcal{O}\left(\frac{1}{N^4}\right)\label{LargENExpTrU}
\end{equation}
where $\langle  \mathscr{T}\mathrm{r} U \rangle_0\equiv \rho_{1}$ (\ref{eq:TrU0}) is the expectation value of $ \mathscr{T}\mathrm{r} U$ evaluated on the saddle point (\ref{eq:bgsol}) and $\langle  \mathscr{T}\mathrm{r} U \rangle_1$ is the subleading contribution. Therefore from (\ref{eq:connectdef}) we see that the subleading contribution to the two point correlator
\begin{equation}
    \langle  \mathscr{T}\mathrm{r} U  \mathscr{T}\mathrm{r} U\rangle = \langle  \mathscr{T}\mathrm{r}~ U\rangle_0^2 + \frac{1}{N^2} \bigg( 2 \langle  \mathscr{T}\mathrm{r} U \rangle_0 \langle  \mathscr{T}\mathrm{r} U \rangle_1 +  \langle  \mathscr{T}\mathrm{r} U  \mathscr{T}\mathrm{r} U\rangle_{c} \bigg)+\mathcal{O}\left( \frac{1}{N^4}\right).
\end{equation}
is the bracketed term multiplying $1/N^2$. It suggests that at the large $N$ perturbation, we can write the Hopf link and unknot invariants (\ref{eq:HUrescale}) as 
\begin{equation}
    \begin{aligned}
\mathscr{V}^{(2,2)}_{\Box,\Box}(S^{3}/\mathbb{Z}_{p},\lambda) & = N^2 \langle  \mathscr{T}\mathrm{r} U\rangle_0^2 + \bigg( 2 \langle  \mathscr{T}\mathrm{r} U \rangle_0 \langle  \mathscr{T}\mathrm{r} U \rangle_1 +  \langle  \mathscr{T}\mathrm{r} U  \mathscr{T}\mathrm{r} U\rangle_{c} \bigg)+\mathcal{O}\left(\frac{1}{N^2} \right)\\
        \mathscr{V}^{(2,1)}_{\Box}(S^{3}/\mathbb{Z}_{p},\lambda) & = N \langle  \mathscr{T}\mathrm{r} U \rangle_0 + \frac{1}{N} \langle  \mathscr{T}\mathrm{r} U \rangle_1 +\mathcal{O}\left( \frac{1}{N^3}\right).
    \end{aligned}
\end{equation}
The subleading correction to Hopf link invariant has two parts. $\langle  \mathscr{T}\mathrm{r} U \rangle_0 \langle  \mathscr{T}\mathrm{r} U \rangle_1$, which being a product of the leading and subleading contributions to the unknot invariant is insensitive to the intertwining  between the two component knots constituting Hopf link. The other contribution $\langle  \mathscr{T}\mathrm{r}U  \mathscr{T}\mathrm{r}U\rangle_{c}$, viz. the connected piece of the two-point correlator, contains information  about how the two unknots are intertwined in a Hopf link when all the representations are in fundamental. That is, it contains the information about the topological nature of the Hopf link. Instead of fundamental, if we consider the Hopf link in some other representations, the structure of the right hand side of (\ref{eq:hopf-unsq}) will be very different. We shall discuss this at the end of this section.

Our next goal is to compute $\langle  \mathscr{T}\mathrm{r}U \rangle_1$ and $\langle  \mathscr{T}\mathrm{r} U  \mathscr{T}\mathrm{r} U\rangle_{c}$ using the methods developed in \cite{eynard2015random, Ambjorn:1992gw}\footnote{The techniques to compute subleading corrections developed in Ref. \cite{eynard2015random, Ambjorn:1992gw} are for hermitian matrix models. Here we develop the similar techniques for unitary matrix models.} and show that the matrix model results are consistent to those obtained from the HOMFLY-PT polynomials.

\subsection{Calculation of \texorpdfstring{$\langle  \mathscr{T}\mathrm{r} U \rangle_1$}{Lg} and \texorpdfstring{$\langle  \mathscr{T}\mathrm{r} U  \mathscr{T}\mathrm{r} U\rangle_{c}$}{Lg}}

Introducing a set of complex variables $z_i = e^{i\theta_i}$, the partition function (\ref{eq:pfU}) can be written as
\begin{equation}\label{eq:PFVz}
    \mathcal{Z}(S^3/\mathbb{Z}_p) = \int \prod_i dz_i \Delta(z_i) e^{-N \sum_{i=1}^N V(z_i)}
\end{equation}
where $\Delta(z_i) = \prod_{i<j}(z_i-z_j)^2$ and the potential $V(z_i)$ is given by
\begin{equation}
    V(z_i) = -\frac{p}{4\pi \lambda} (\ln z_i)^2 + \ln z_i-\frac{\pi p}{12} .
\end{equation}
In the large $N$ limit with the continuous variables $x=i/N$, $z_i \rightarrow z(x)$ and $\sum_i = N \int_0^1 dx$, we can write
\begin{eqnarray}
    -N \sum_{i=1}^N V(z_i) = -N^2 \int_{0}^1 dx V(z(x))
\end{eqnarray}
where, 
\begin{equation}\label{eq:Vz}
    V(z) = -\frac{p}{4\pi \lambda} (\ln z)^2 + \ln z -\frac{\pi p}{12}.
\end{equation}
Following (\ref{eq:evdensity}) we define the eigenvalue density in the complex $z$-plane as
\begin{equation}\label{eq:evdenc}
    \rho(z) = \left\langle\frac{1}{N} \sum_{i=1}^N \delta(z-z_i) \right\rangle.
\end{equation}
In the large $N$ limit the above density satisfies the saddle point equation:
\begin{equation}
    2 \dashint dw \frac{\rho(w)}{z-w} = V'(z) .
\end{equation}

\subsubsection{One-loop correlator}
We introduce the one-loop correlator (also known as resolvent) $R(z)$ as
\begin{equation}
    R(z) = \frac{1}{N} \left \langle \Tr \frac{1}{z-U}\right\rangle\equiv \frac{1}{N}\sum_{m=0}^{\infty}z^{-1-m}\langle \Tr U^{m}\rangle.
\end{equation}
Since $\langle \Tr U^m\rangle/N \leq 1$ for any $m$, $R(z)$ is an analytic function in complex $z$ plane inside and outside the circle $|z| = 1$. In diagonal basis eigenvalues of $U$ lies on $|z|=1$ circle. Hence the function $R(z)$ has poles at the eigenvalues of $U$. In the large $N$ limit, the eigenvalues get accumulated and form a distribution on disconnected finite arcs on the unit circle. As a result $R(z)$ develops branch cut singularities on these arcs. Using (\ref{eq:evdenc}) we can write
\begin{equation}
    R(z) = \int dw \frac{\rho(w)}{z-w}.\label{ResolventUsingrho}
\end{equation}
Using the Sokhotski-Plemelj theorem 
\begin{align}
    \lim_{\epsilon \rightarrow 0} \int dx \frac{f(x)}{x_0 -x\mp i \epsilon} &= \dashint dx \frac{f(x)}{x_0-x} \pm i \pi f(x_0)
\end{align}
 we find that the resolvent satisfies
\begin{equation}\label{eq:Rzcpndition+}
   \lim_{\epsilon \rightarrow 0} \big( R(z + i \epsilon) + R(z- i \epsilon) \big)= V'(z) 
\end{equation}
and
\begin{equation}\label{eq:Rzcpndition-}
  \lim_{\epsilon\rightarrow 0}  \big( R(z + i \epsilon) - R(z- i \epsilon)\big) = -2\pi i \rho(z).
\end{equation}

Moreover $R(z)$ has the following asymptotic $(z \rightarrow \infty)$ structure
\begin{equation}
    \label{eq:Rzasym}
    R(z\rightarrow \infty) = \frac{1}{z} + \frac{1}{z^2} \langle \mathscr{T}\mathrm{r}~ U\rangle + \frac{1}{z^3} \langle \mathscr{T}\mathrm{r}~ U^2\rangle + \mathcal{O}(1/z^4).
\end{equation}
Thus $R(z)$ is the generating function for $\langle \mathscr{T}\mathrm{r}~ U^m\rangle$ for different $m>0$. The resolvent $R(z)$ admits a perturbative large $N$ expansion
\begin{equation}
    R(z) = R_0(z) + \frac{1}{N^2} R_1(z) + \cdots
\end{equation}
The leading contribution of the resolvent, $R_0(z)$ can be obtained by solving the Dyson-Schwinger equation \cite{Friedan:1980tu, Dutta:2016byx} in the $N \rightarrow \infty$ limit. $R_0(z)$ also has a branch cut on the unit circle. The eigenvalue density $\rho(z)$  can be determined from the discontinuity of $R_0(z)$ on the branch cut. The higher genus contributions $R_1(z)$, $R_2(z)$ \textit{et cetra} then can be calculated using loop equations order by order \cite{Ambjorn:1992gw}. For a given potential $V(w)$, if the eigenvalue density has support between $a$ and $b$ on the unit circle then $R_0(z)$ can be expressed as \cite{Migdal:1983qrz}
\begin{equation}
      R_0(z) = \frac{1}{2} \oint_{\mathcal{C}} \frac{dw}{2\pi i} \frac{V'(w)}{z-w} \sqrt{\frac{(z-a)(z-b)}{(w-a)(w-b)}}\label{eqn:R0(Z)}
\end{equation}
where $\mathcal{C}$ is a counter-clockwise contour around the branch cut between $a$ and $b$. The spectral edges $(a,b)$ can be found using the properties of the asymptotic expansion of $R_{0}(z)$. We can do the contour integral (\ref{eqn:R0(Z)}) for the potential $V(w)$ given in equation (\ref{eq:Vz}). {Since the potential (\ref{eq:weight}) is an even function of $\theta$, the eigenvalue density is symmetric about $\theta =0$.} As a result, in a complex $z$-plane the branch cuts of $R(z)$ are symmetrically located on the unit circle about the real axis. Hence we can take
\begin{equation}
a = b^{-1} = e^{i \theta}. 
\end{equation}
The final answer is given by,
\begin{eqnarray}
    R_0(z) = \frac{1}{2 z}-\frac{p}{4 \pi  \lambda  z} \log \left[\frac{ z^2 \left(\cos \theta  - z +\sqrt{z^2 - 2 \ z\ 
   \cos \theta + 1}\right)}{1 - z \ \cos \theta + \sqrt{z^2-2\ z \ \cos
   \theta+1}}\right].
\end{eqnarray}
It is easy to check that $R_0(z)$ has a branch cut singularity between $\theta = \pm 2 \sec^{-1}e^{\pi \lambda/p}$ on the unit circle in the complex $z$ plane. This implies the spectral edges are given by 
\begin{align}
a=b^{-1}=\exp\big({2i \sec^{-1}e^{\pi \lambda /p}}\big).    \label{SpectRalEdg}
\end{align}
Using an expression analogous to (\ref{eq:Rzcpndition-}), we can calculate the discontinuity of $R_{0}(z)$ and find the leading eigenvalue density. The answer is  given in (\ref{eq:bgsol}). The coefficient of $1/z^2$ in the asymptotic of $R_0(z)$ can be computed and that matches with leading contribution of the expectation value of $ \mathscr{T}\mathrm{r}~ U$ given in (\ref{eq:TrU0}) for $p=1$.

Our next goal is to compute $R_1(z)$.  The coefficient of $1/z^2$ in the asymptotic expansion of $R_1(z)$ gives $1/N^2$ correction of $\langle \mathscr{T}\mathrm{r} U\rangle$ viz. $\langle \mathscr{T}\mathrm{r} U\rangle_1$. We follow the prescription given in \cite{Ambjorn:1992gw} to write $R_1(z)$ in terms of a series of functions. We shall not elaborate the method in this paper. $R_1(z)$ is given by
\begin{align}
    R_1(z) & = \frac{1}{8(a-b)}\left( \psi^{(1)}(z) - \chi^{(1)}(z) \right) +\frac{1}{16} \left( \psi^{(2)}(z) + \chi^{(2)}(z) \right)\label{eqn:R1(z)}
\end{align}
where
\begin{equation}
\begin{aligned}
    \chi^{(1)}(z) & = \frac{1}{M_1} \Phi_a^{(1)}(z), \\ 
    \psi^{(1)}(z) & = \frac{1}{J_1} \Phi_b^{(1)}(z), \\
    \chi^{(2)}(z) & = \frac{1}{M_1} \left(\Phi_a^{(2)}(z) - \chi^{(1)}(z) M_2\right), \\
    \psi^{(2)}(z) & = \frac{1}{J_1} \left( \Phi^{(2)}_b(z) - \psi^{(1)}(z) J_2 \right) .
    \end{aligned}
\end{equation}
The functions $\Phi^{(n)}_a(z)$ and $\Phi^{(n)}_b(z)$ are given by
\begin{equation}
    \begin{aligned}
        \Phi^{(n)}_a(z) & = \frac{1}{(z-a)^n \sqrt{(z-a)(z-b)}},\\
        \Phi^{(n)}_b(z) & = \frac{1}{(z-b)^n \sqrt{(z-a)(z-b)}}.
    \end{aligned}
\end{equation}
$M_n$ and $J_n$ are given by the following contour integrals
\begin{equation}\label{eq:MnJn}
    \begin{aligned}
        M_n & = \oint_{\mathcal{C}} \frac{d\omega}{2\pi i } \frac{V'(\omega)}{(\omega-a)^n\sqrt{(\omega-a)(\omega-b)}}, \\
        J_n & = \oint_{\mathcal{C}} \frac{d\omega}{2\pi i } \frac{V'(\omega)}{(\omega-b)^n\sqrt{(\omega-a)(\omega-b)}}.
    \end{aligned}
\end{equation}
Evaluating the contour integrals (\ref{eq:MnJn}) we find
\begin{equation}
    \begin{aligned}
        M_1 & = \frac{4 A}{a^{3/2} \left(\sqrt{a}+\sqrt{b}\right)}, \qquad \quad \ \ 
        J_1  = \frac{4 A}{b^{3/2} \left(\sqrt{a}+\sqrt{b}\right)} \\
        M_2 & = -\frac{4 A \left(5 \sqrt{a}+4 \sqrt{b}\right)}{3
   a^{5/2} \left(\sqrt{a}+\sqrt{b}\right)^2}, \qquad J_2  = -\frac{4 A \left(4 \sqrt{a}+5 \sqrt{b}\right)}{3
   b^{5/2} \left(\sqrt{a}+\sqrt{b}\right)^2},
    \end{aligned}
\end{equation}
where $A=-\frac{p}{4\pi \lambda}$. Plugging all these expressions in (\ref{eqn:R1(z)}) we finally arrive at
\begin{eqnarray}
    R_1(z) = - \frac{\pi  \lambda  (1 + z) (1 - 8 z \cos \theta - 6 z + z^2) \ \sin ^2\frac{\theta
   }{2} }{12 p
   \left(1 - 2 z \cos \theta  + z^2 \right)^{5/2}}.
\end{eqnarray}
We expand $R_1(z)$ asymptotically to obtain
\begin{equation}
    R_1(z) = - \frac{\pi  \lambda \ \sin ^2\frac{\theta
   }{2}}{12 \ p \ z^2}  + \frac{\pi  \lambda\   \sin ^2\frac{\theta
   }{2} (3 \cos \theta +5)}{12 \ p\ z^3} +\mathcal{O}\left( \frac{1}{z^4}\right).
\end{equation}
Thus from the asymptotic expansion we collect the coefficient of $1/z^2$ and get
\begin{equation}
    \langle  \mathscr{T}\mathrm{r} U\rangle_1 = - \frac{\pi  \lambda \ \sin ^2\frac{\theta
   }{2}}{12 \ p} = - \frac{\pi  \lambda  \left(1-e^{-\frac{2 \pi  \lambda
   }{p}}\right)}{12 p} .\label{SubLUnMatM}
\end{equation}
Following the prescription (\ref{eq:MaPPing}) we find 
\begin{equation}
    \langle  \mathscr{T}\mathrm{r} U\rangle_1 = \frac{i \pi  \lambda  \left(1-e^{2i  \pi  \lambda
   }\right)}{12} .\label{SubLeAdUnknot}
\end{equation}
The result (\ref{SubLeAdUnknot}) matches exactly with the subleading contribution of the vertical framing $U(N)$ invariant of an unknot $\mathcal{W}_{\Box}^{(2,1)}(S^3,\lambda)$ carrying fundamental representation on its component.
\subsubsection{Two-loop correlator}

We next compute the \emph{leading} connected piece of two point function $\langle  \mathscr{T}\mathrm{r}U  \mathscr{T}\mathrm{r}U \rangle$. We define two-loop correlator as
\begin{equation}
    R(z,w) = \frac{1}{N^2} \left\langle \Tr \frac{1}{z-U} \Tr\frac{1}{w-U}\right\rangle.
\end{equation}
In the large $N$ limit two-loop correlator is equal to the product of two one-loop correlators $R(z)$ and $R(w)$. There is a connected subleading piece, denoted by $R^{(c)}(z,w)$ and suppressed by a factor of $N^2$
\begin{equation}
    R(z,w) = R(z) R(w) + \frac{1}{N^2} R^{(c)}(z,w).
\end{equation}
Our goal is to find the connected piece of the two-point correlator. For the partition function (\ref{eq:PFVz}), the eigenvalue density can be written as
\begin{equation}
    \rho(z) = \left\langle\frac{1}{N} \sum_{i=1}^N \delta(z-z_i) \right\rangle = - \frac{1}{N^2} \frac{\delta}{\delta V(z)} \log \mathcal{Z}.
\end{equation}
Here we have used the fact that $\frac{\delta V(x)}{\delta V(y)} = \delta(x-y)$. Repeating the similar analysis we can write
\begin{equation}
    \frac{\delta \rho(z)}{\delta V(w)} = -N^2 \left(  \rho(z, w)  -   \rho(z)  \rho(w) \right)
\end{equation}
where
\begin{equation}\label{eq:rho2}
    \rho(z, w) = \frac{1}{N^2} \left\langle \sum_{i,j=1}^N \delta(z - z_i) \delta(w - z_i) \right\rangle.
\end{equation}
In the large $N$ limit, $ \rho(z, w) $ is also decomposed into a product of two eigenvalue densities $ \rho(z)$ and $  \rho(w) $. The connected subleading contribution $ \rho_c(z, w)$ is suppressed by $1/N^2$ factor:
\begin{equation}\label{eq:rho2c}
    \rho(z, w)  = \rho(z)  \rho(w)  + \frac{1}{N^2} \rho_c(z, w).
\end{equation}
Hence we find
\begin{equation}
    \frac{\delta \rho(z)}{\delta V(w)} = - \rho_c(z, w). 
\end{equation}
We know that using the definition of eigenvalue density, the one-loop correlator $R(z)$ can be expressed as (\ref{ResolventUsingrho}). After a little algebra we find that
\begin{equation}
    \int \frac{dv}{w-v} \frac{\delta R(z)}{\delta V(v)} = - \int du \int dv \frac{\rho_c(u,v)}{(z-u)(w-v)}.
\end{equation}
Using the definition (\ref{eq:rho2}) we can express two-point correlator as
\begin{equation}
    R(z,w) = \int du \int dv \frac{\rho(u,v)}{(z-u)(w-v)}.
\end{equation}
Likewise for the connected components of $R(z,w)$ and $\rho(u,v)$ we can write
\begin{equation}
    R^{(c)}(z,w) = \int du \int dv \frac{\rho_c(u,v)}{(z-u)(w-v)}.
\end{equation}
Thus we have
\begin{equation}
    \int \frac{dv}{w-v} \frac{\delta R(z)}{\delta V(v)} = - R^{(c)}(z,w).
\end{equation}
Using the analytic properties (\ref{eq:Rzcpndition+}) of $R(z)$ about the cut we find that
\begin{equation}
    R^{(c)}(z + i \epsilon, w) + R^{(c)}(z - i \epsilon, w) = -\frac{1}{(z-w)^2} .
\end{equation}
Since $R(z,w)$ is symmetric in $z$ and $w$ similar property holds for the $w$ variable also. The connected piece of the resolvent, $R^{(c)}(z,w)$ being discontinuous about the cut, $R^{(c)}(z,w) + \frac{1}{2(z-w)^2}$ must change sign as we cross the cut for both the arguments. Therefore we take an ansatz for $R^{(c)}(z,w)$ as follows
\begin{equation}
    R^{(c)}(z,w) = -\frac{1}{2(z-w)^2} \left(1 - \frac{Q(z,w)}{\sqrt{\sigma(z)} \sqrt{\sigma(w)}} \right), \quad \text{where} \quad \sigma(z) = (z-a)(z-b).
\end{equation}
$Q(z,w)$ is symmetric meromorphic in its arguments. $Q(z,w)$ can be found for a one-gap solution from the properties of $R^{(c)}(z,w)$. From the definition of $R(z,w)$ we see that it is regular as $z \rightarrow w$. Expanding $R^{(c)}(z,w)$ near $w=z$ we arrive at
\begin{equation}
    R^{(c)}(z,w) = -\frac{1}{2(z-w)^2} \left(1 - \frac{Q(z,z)}{\sigma(z)} - \frac{(w-z)}{\sigma(z)} \left( \partial_w Q(z,w) - \frac{\sigma'(w)Q(z,w)}{2\sigma(z)}\right)\bigg|_{w=z}\right).
\end{equation}
Therefore to make $ R^{(c)}(z,w)$ finite at $z=w$ we have
\begin{align}
    \lim_{w \rightarrow z} Q(z,w) &= \sigma(z)~~\text{and}~~\lim_{w\rightarrow z} \partial_w Q(z,w) = \frac{\sigma'(z)}{2}.
\end{align} $Q(z,w)$ must be regular at spectral edges hence $Q(z,w)$ must be a polynomial. Since $R(z,z)$ goes as $1/z^2$ asymptotically, for one-cut solution $Q(z,z)$ must be a polynomial of order 2. The above conditions can uniquely determine $Q$ for one-cut solution and it is given by
 \begin{equation}
     Q(z,w) = z w - \frac{1}{2}(a+b)(z+w) + ab.
 \end{equation}
The leading contribution of $\left\langle  \mathscr{T}\mathrm{r}U  \mathscr{T}\mathrm{r} U\right\rangle_{c}$ can be obtained from $\frac{1}{z^2 w^2}$ coefficient of $R^{(c)}(z,w)$ in its asymptotic expansion. After a careful computation we find that
\begin{equation}
 \boldsymbol{\ell}.\boldsymbol{c}.\left[   \langle  \mathscr{T}\mathrm{r} U   \mathscr{T}\mathrm{r} U \rangle_c \right] = \frac{(a-b)^2}{16} = e^{-\frac{2\pi \lambda}{p}}\left( e^{-\frac{2\pi \lambda}{p}}-1\right),\label{TrUTrUMtM}
\end{equation}
where we have used the expressions of the spectral edges (\ref{SpectRalEdg}). With the prescription (\ref{eq:MaPPing}) of analytic continuation $p\rightarrow ip$, followed by setting $p=1$ we arrive at 
\begin{equation}
 \boldsymbol{\ell}.\boldsymbol{c}. \left[  \langle  \mathscr{T}\mathrm{r} U  \mathscr{T}\mathrm{r} U \rangle_c\right] = e^{2 i \pi \lambda} (e^{2 i \pi \lambda} -1 ).\label{ConnecCorr}
\end{equation}
The result (\ref{ConnecCorr}) matches exactly with the following result from the HOMFLY-PT polynomials:
\begin{align}
\mathcal{W}_{\Box,\Box}^{[(2,2),f=\{1,1\}]}(S^3,\lambda)-\left(\mathcal{W}_{\Box}^{(2,1)} (S^3,\lambda)\right)^2 .\label{Hopf11modUnknot1}
\end{align}
This suggests that $\left\langle \mathscr{T}\mathrm{r}U\mathscr{T}\mathrm{r}U\right\rangle_{c}$ evaluated in (\ref{TrUTrUMtM}) is exact in $N$. Any further large $N$ corrections of $\left\langle \mathscr{T}\mathrm{r}U \mathscr{T}\mathrm{r}U\right\rangle_{c}$ from the matrix model approach will vanish. Thus we have 
\begin{align}
    \left\langle \mathscr{T}\mathrm{r}U\mathscr{T}\mathrm{r}U\right\rangle_{c}&=e^{-\frac{2\pi \lambda}{p}}\left( e^{-\frac{2\pi \lambda}{p}}-1\right).
\end{align}

\subsection{Knot/link invariants and connected correlators}
\label{sec:knotandcorrel}
So far we have confined ourselves to Hopf link invariant with fundamental representation placed on its components. Now we look at the connected piece of two point correlator carrying some low dimensional representations apart from the fundamental. Using the fact that Schur polynomials for any arbitrary representation can be written as sum over different powers of $\Tr U^m$ (\ref{eq:chaexp}), we can express the large $N$ results of any (torus) knot or link invariant modulo its leading contribution in terms of connected correlation functions of $\Tr U^m$. For example, consider the unknot invariant in representation $\ydiagram[]{2}$, i.e. $\mathscr{V}_{\tiny{\ydiagram[]{2}}}^{(2,1)}(S^3/\mathbb{Z}_{p},\lambda)$. Using (\ref{LeadingContofUnknot}) the leading contribution can be read off to be
\begin{align}
\boldsymbol{\ell}.\boldsymbol{c}.\left[\mathscr{V}^{(2,1)}_{\tiny{\ydiagram[]{2}}}(S^3/\mathbb{Z}_{p},\lambda)\right]&=\frac{1}{2} \left(\boldsymbol{\ell}.\boldsymbol{c}.\left[\mathscr{V}_{\tiny{\ydiagram[]{1}}}^{(2,1)}(S^3/\mathbb{Z}_{p},\lambda)\right]\right)^2.
\end{align}
 We know that  (\ref{L,LInvS^3/Z_p}, \ref{eq:chaexp})
 \begin{align}
    \mathscr{V}_{\tiny{\ydiagram[]{2}}}^{(2,1)}(S^3/\mathbb{Z}_{p},k)& = \langle \mathbf{s}_{\tiny{\ydiagram[]{2}}}\rangle ,~~~
  \text{where}~~~  \mathbf{s}_{\tiny{\ydiagram[]{2}}} =\frac{N}{2} \left( \mathscr{T}\mathrm{r} U^2 \right)+ \frac{N^2}{2}  \left(\mathscr{T}\mathrm{r} U \mathscr{T}\mathrm{r} U \right).
 \end{align}
Hence we have  
\begin{align}
\mathscr{V}^{(2,1)}_{\tiny{\ydiagram[]{2}}}(S^3/\mathbb{Z}_{p},k) - \frac{1}{2} \left(\mathscr{V}_{\tiny{\ydiagram[]{1}}}^{(2,1)}(S^3/\mathbb{Z}_{p},k)\right)^2 &= \frac{N}{2}\left\langle \mathscr{T}\mathrm{r} U^2 \right\rangle +  \frac{1}{2}\left\langle \mathscr{T}\mathrm{r} U \mathscr{T}\mathrm{r} U \right\rangle_c.\label{Unknot2BOX}
\end{align}
This expression can also seen to be consistent using cabling prescription \cite{ramadevi2000computation}
and Frobenius relation in group theory \cite{labastida2001knots,marino2001framed,ramadevi2012reformulated}.
In a similar fashion $\mathscr{V}_{\tiny{\ydiagram[]{1,1}}}^{(2,1)}(S^3/\mathbb{Z}_{p},k)$ admits a relation
\begin{align}
\mathscr{V}_{\tiny{\ydiagram[]{1,1}}}^{(2,1)}(S^3/\mathbb{Z}_{p},k) - \frac{1}{2} \left(\mathscr{V}_{\tiny{\ydiagram[]{1}}}^{(2,1)}(S^3/\mathbb{Z}_{p},k)\right)^2 = -\frac{N}{2}\langle \mathscr{T}\mathrm{r} U^2 \rangle + \frac{1}{2} \langle \mathscr{T}\mathrm{r} U \mathscr{T}\mathrm{r} U \rangle_c.\label{Unknot11BOX}
\end{align}
Thus we see that when we change the representations, the unknot invariants (modulo leading contribution) have different expansions in terms of correlation functions of $\mathscr{T}\mathrm{r}U^{m}$ operators.

Similar analysis can be done for a two component Hopf link whose invariant $\mathscr{V}^{(2,2)}_{\tiny{\ydiagram[]{2},\ydiagram[]{1}}}(S^{3}/\mathbb{Z}_{p},k) = \langle \mathbf{s}_{\tiny{\ydiagram[]{2}}}\mathbf{s}_{\tiny{\ydiagram[]{1}}}\rangle$ satisfies:
\begin{align}
\label{eq:hopf21}
\mathscr{V}^{(2,2)}_{\tiny{\ydiagram[]{2},\ydiagram[]{1}}}(S^{3}/\mathbb{Z}_{p},k) - \mathscr{V}_{\tiny{\ydiagram[]{2}}}^{(2,1)} (S^3/\mathbb{Z}_{p},k)\mathscr{V}_{\tiny{\ydiagram[]{1}}}^{(2,1)}(S^3/\mathbb{Z}_{p},k) &= N \langle \mathscr{T}\mathrm{r}U \rangle \langle \mathscr{T}\mathrm{r}U \mathscr{T}\mathrm{r}U \rangle_c \nonumber\\
+ \frac{1}{2}\langle \mathscr{T}&\mathrm{r}U \mathscr{T}\mathrm{r}U \mathscr{T}\mathrm{r}U \rangle_c + \frac{1}{2}\langle \mathscr{T}\mathrm{r}U \mathscr{T}\mathrm{r}U^2 \rangle_c.
\end{align}
In fact any $n$ point correlator can be expressed in terms of the connected correlators as
\begin{equation}\label{eq:conndef}
\begin{aligned}
    \langle \underbrace{\mathscr{T}\mathrm{r}U \cdots \mathscr{T}\mathrm{r}U}_{n\ \text{times}}&\rangle = \langle \mathscr{T}\mathrm{r}U\rangle^n  + \frac{^nC_2}{N^2} \langle \mathscr{T}\mathrm{r}U \rangle^{n-2} \langle \mathscr{T}\mathrm{r}U\mathscr{T}\mathrm{r}U\rangle_c \\
     &~~~~+ \frac{^nC_3}{N^3} \langle \mathscr{T}\mathrm{r}U \rangle^{n-3} \langle \mathscr{T}\mathrm{r}U\mathscr{T}\mathrm{r}U \mathscr{T}\mathrm{r}U\rangle_c \\
     &~+\frac{^nC_4}{N^4}\langle \mathscr{T}\mathrm{r}U \rangle^{n-4} \left( ^4C_2\langle \mathscr{T}\mathrm{r}U\mathscr{T}\mathrm{r}U \rangle_c^2 + \langle \mathscr{T}\mathrm{r}U \mathscr{T}\mathrm{r}U \mathscr{T}\mathrm{r}U \mathscr{T}\mathrm{r}U\rangle_c  \right)\\
     +&\frac{^nC_5}{N^5} \langle \mathscr{T}\mathrm{r}U\rangle^{n-5}\left(^5C_2 \langle \mathscr{T}\mathrm{r}U \mathscr{T}\mathrm{r}U\rangle_{c}\langle \mathscr{T}\mathrm{r}U \mathscr{T}\mathrm{r}U\mathscr{T}\mathrm{r}U\rangle_{c}+\langle \left(\mathscr{T}\mathrm{r}U \right)^5\rangle_{c}\right)\\
    +\frac{^nC_6}{N^6} &\langle \mathscr{T}\mathrm{r}U\rangle^{n-6}\left(^6C_2 \langle \left(\mathscr{T}\mathrm{r}U \right)^2 \rangle_{c}\langle \left(\mathscr{T}\mathrm{r}U\right)^4\rangle_{c} + ^6C_3 \langle \left(\mathscr{T}\mathrm{r}U \right)^3\rangle^2_{c} +\langle \left(\mathscr{T}\mathrm{r}U \right)^6\rangle_{c}\right)\\
    +\cdots~~~ &.
    \end{aligned}
\end{equation}
In order to validate the result with the colored HOMFLY-PT polynomial we compute the leading contribution of (\ref{eq:hopf21}) in the matrix model side. It is given by (\ref{eq:TrU0},\ref{TrUTrUMtM})
\begin{align}
\boldsymbol{\ell}.\boldsymbol{c}.&\left[\mathscr{V}^{(2,2)}_{\tiny{\ydiagram[]{2},\ydiagram[]{1}}}(S^{3}/\mathbb{Z}_{p},\lambda) - \mathscr{V}_{\tiny{\ydiagram[]{2}}}^{(2,1)}(S^3/\mathbb{Z}_{p},\lambda) \mathscr{V}_{\tiny{\ydiagram[]{1}}}^{(2,1)}(S^3/\mathbb{Z}_{p},\lambda)\right]= N \langle \mathscr{T}\mathrm{r}U \rangle_{0} \langle \mathscr{T}\mathrm{r}U \mathscr{T}\mathrm{r}U \rangle_c\nonumber\\
&~~~~~~~~~~~~~~~~~~~~~~~~~~~~~~~~~~~~~~~~~~~~~~~~~~=Np\left(\frac{1-e^{-\frac{2\pi \lambda}{p}}}{2\pi \lambda}\right)e^{-\frac{2\pi \lambda}{p}}\left(e^{-\frac{2\pi \lambda}{p}}-1 \right).\label{eq:lcHopf2,1}
\end{align}
Performing the analytic continuation $p\rightarrow ip$ followed by setting $p=1$ (\ref{eq:MaPPing}) the above expression becomes identical to the one obtained from the corresponding colored HOMFLY-PT polynomials:
\begin{align}
\boldsymbol{\ell}.\boldsymbol{c}.&\left[\mathcal{W}^{[(2,2);f=\{1,1\}]}_{\tiny{\ydiagram[]{2},\ydiagram[]{1}}}(S^{3},\lambda) - \mathcal{W}_{\tiny{\ydiagram[]{2}}}^{(2,1)}(S^3,\lambda) \mathcal{W}_{\tiny{\ydiagram[]{1}}}^{(2,1)}(S^3,\lambda)\right]=  \frac{2 i N}{\pi \lambda} e^{4\pi i \lambda} \sin^2 \pi \lambda .
\end{align} We provide one more example before concluding this section.
\begin{equation}
\begin{aligned}\label{eq:hopf22}
\mathscr{V}^{(2,2)}_{\tiny{\ydiagram[]{2},\ydiagram[]{2}}}(S^{3}/\mathbb{Z}_{p},k) - \left(\mathscr{V}_{\tiny{\ydiagram[]{2}}}^{(2,1)}(S^3/\mathbb{Z}_{p},k)\right)^2  = \ \ & N^2 \langle \mathscr{T}\mathrm{r}U\rangle^2 \langle \mathscr{T}\mathrm{r}U \mathscr{T}\mathrm{r}U\rangle_c \\
         +  N \big( \langle \mathscr{T}\mathrm{r}U& \rangle \langle \mathscr{T}\mathrm{r}U\mathscr{T}\mathrm{r}U\mathscr{T}\mathrm{r}U \rangle_c + \langle \mathscr{T}\mathrm{r}U \rangle \langle \mathscr{T}\mathrm{r}U\mathscr{T}\mathrm{r}U^2\rangle_c \big) \\
         +  \frac{5}{4} \langle \mathscr{T}&\mathrm{r}U \mathscr{T}\mathrm{r}U\rangle_c^2 + \frac{1}{4} \langle \mathscr{T}\mathrm{r}U \mathscr{T}\mathrm{r}U \mathscr{T}\mathrm{r}U \mathscr{T}\mathrm{r}U\rangle_c \\
         +  \frac{1}{4}&\langle (\mathscr{T}\mathrm{r}U^2)^2 \rangle_c +\frac{1}{2} \langle (\mathscr{T}\mathrm{r}U)^2 \mathscr{T}\mathrm{r}U^2 \rangle_c.
\end{aligned}
\end{equation}
Again we can compute the leading contribution of (\ref{eq:hopf22}) in the double scaling limit using the matrix model results (\ref{eq:TrU0},\ref{TrUTrUMtM}):
\begin{align}    \boldsymbol{\ell}.\boldsymbol{c}.&\left[\mathscr{V}^{(2,2)}_{\tiny{\ydiagram[]{2},\ydiagram[]{2}}}(S^{3}/\mathbb{Z}_{p},\lambda) - \left(\mathscr{V}_{\tiny{\ydiagram[]{2}}}^{(2,1)}(S^3/\mathbb{Z}_{p},\lambda)\right)^2 \right]=N^{2}\left\langle \mathscr{T}\mathrm{r}U\right\rangle_{0}^{2}\left\langle \mathscr{T}\mathrm{r}U\mathscr{T}\mathrm{r}U\right\rangle_{c}\nonumber\\
&~~~~~~~~~~~~~~~~~~~~~~~~~~~~~~~~~~~~~~~~~~~~~~=N^2p^2\left(\frac{1-e^{-\frac{2\pi \lambda}{p}}}{2\pi \lambda}\right)^2e^{-\frac{2\pi \lambda}{p}}\left(e^{-\frac{2\pi \lambda}{p}}-1 \right).\label{eq:lcHopf2,2}
\end{align}
Now we analytically  continue $p\rightarrow ip$ and set $p=1$ (\ref{eq:MaPPing}) to get 
\begin{equation}
\boldsymbol{\ell}.\boldsymbol{c}.\left[\mathcal{W}^{[(2,2);f=\{1,1\}]}_{\tiny{\ydiagram[]{2},\ydiagram[]{2}}}(S^{3},\lambda) - \left(\mathcal{W}_{\tiny{\ydiagram[]{2}}}^{(2,1)}(S^3,\lambda) \right)^2\right]= \frac{2 i N^2}{\pi^2 \lambda^2} e^{5\pi i \lambda} \sin^3 \pi \lambda .
\end{equation}
We see from equations (\ref{eq:hopf-unsq}, \ref{eq:hopf21}, \ref{eq:hopf22}) that as we change the representations associated with two of the components of Hopf link, the expansion in terms of correlation functions also get modified. Therefore any unknot or Hopf link invariant (modulo the leading contribution) with arbitrary representations placed on the components can be uniquely expressed in terms of different connected correlation functions. For example, from equations (\ref{Unknot2BOX}) and (\ref{Unknot11BOX}) we see that unknots in representations $\ydiagram[]{2}$ and $\ydiagram[]{1,1}$ are expressed in terms of $\langle \mathscr{T}\mathrm{r}U \rangle$ and $\langle \mathscr{T}\mathrm{r}U \mathscr{T}\mathrm{r}U\rangle_c$. However they differ by the sign of $\langle \mathscr{T}\mathrm{r}U \rangle$ term. 
Using the character expansion for Schur polynomial (\ref{eq:chaexp}) and equation (\ref{eq:conndef}) one can similarly express any $(L,L)$ torus link invariant uniquely in terms of connected correlation functions. 

It is also to be noted that the evaluation of the corresponding leading contributions as in equations (\ref{eq:lcHopf2,1}, \ref{eq:hopf22}) using the matrix model results (\ref{eq:TrU0}, \ref{TrUTrUMtM}) was possible because we have placed low dimensional representations on the two component knots of the Hopf link. If we put arbitrary large representations it will be difficult to evaluate the explicit expressions of leading contributions with the matrix model techniques. This is because the results  (\ref{eq:TrU0}, \ref{TrUTrUMtM}) involve the density distribution which extremises the partition function of the manifold and any link carrying large representation will back react on that solution. 

Moreover, for any arbitrary $L$ component torus link the computation of the connected correlator seems technically challenging from the matrix model side, even with low dimensional representations placed on the component knots. Because such invariants will involve higher point connected correlators, which are difficult to handle. Nonetheless, for low representations the colored HOMFLY-PT polynomial (\ref{LAlp,LBeTa}) allows us to obtain the subleading contribution for any $L>2$. For example the polynomials tabulated in Table \ref{Tableof3,3lc} will help us to get the subleading contribution to the $(3,3)$ torus link invariant.

\section{Conclusion}
\label{sec:conclu}
In this paper, we have discussed the torus link invariants in $U(N)$ Chern-Simons theory embedded inside three manifold $S^3$ as well as $S^3/\mathbb{Z}_{p}$. In the double scaling limit (\ref{eq:dslimit1}), it has been shown that any $(L\alpha,L\beta)$ torus link invariant in $S^3$ (\ref{LAlp,LBeTa}) can be expressed in terms of $(L,L)$ torus link invariant, here $\alpha$ and $\beta$ are coprime to each other. Our result is depicted in equation (\ref{LAlphLbetaL,L}). Using the explicit expression of the $(L\alpha,L\beta)$ invariant  (\ref{LAlp,LBeTa}) we can evaluate the polynomial for some low dimensional representations placed on the $L$ component knots of the link (e.g. see Table \ref{Tableof3,3lc}). Consequently, in the large $N$ limit, we can deduce their leading and subleading contributions. However, for links carrying higher dimensional representations this procedure seems practically challenging. Matrix model techniques seem to provide some handle of such difficulties at certain instances. For example, when the representations placed on the component knots of any $(L,L)$ torus link (\ref{L,LInvS^3/Z_p}) is large (i.e. comparable to rank $N$ of the gauge group in the limit $N\rightarrow \infty$) but completely symmetric we can determine the leading contribution (\ref{eq:barH}) to the invariant by using the eigenvalue density (\ref{eq:bgsol}) that dominates partition function (\ref{eq:partitionfunction}) of the manifold $S^3/\mathbb{Z}_{p}$. Moreover, for low dimensional representations, using the aforementioned density distribution, we have written down an analytic expression of the leading contribution of any $(L,L)$ torus link invariant (\ref{l.c(L,L)}). The leading contribution of the torus link invariant being proportional to the leading contribution of the unknot invariant, it does not capture the intertwining information of the component knots constituting the link. To capture such an information the study of connected piece of the correlators in Chern-Simons theory become necessary. From the matrix model approach, we analyse in detail the connected piece of a two point correlator (\ref{eq:connectdef}), which is basically the subleading contribution to the Hopf link invariant when both the component knots carry fundamental representations (\ref{eq:hopf-unsq}). Our detailed analysis lead to the result in equation (\ref{TrUTrUMtM}) and it matches exactly with the one obtained using the HOMFLY-PT polynomials (\ref{ConnecCorr}). We also obtain the subleading contribution to the unknot invariant carrying fundamental representation (\ref{SubLUnMatM}), using the techniques of matrix model. It is interesting to see that if we place low dimensional representations other than fundamental then the Hopf link or unknot invariant modulo the corresponding leading contribution take different structures expressed in terms of correlators of $\mathscr{T}\mathrm{r}U^m$  operators as shown in equations (\ref{Unknot2BOX}, \ref{Unknot11BOX}, \ref{eq:hopf21}, \ref{eq:hopf22}). As a validation, we specifically evaluate the leading contribution of (\ref{eq:hopf21}) and (\ref{eq:hopf22}) using our matrix model computations (\ref{eq:TrU0}, \ref{TrUTrUMtM}) and find the result to agree with the respective colored-HOMFLY-PT polynomials.

Note that, in the double scaling limit Chern-Simons theory admits a third order phase transition similar to Douglas-Kazakov phase transition \cite{Chakraborty:2021oyq}. This means the integrable representation (\ref{eq:bgsol}) that dominates the partition function (\ref{eq:partitionfunction})  changes at some critical value of $\lambda$. As a result the correlation functions when evaluated on the leading solution must change its behaviour at the critical value. However, the origin of such behaviour is not clear from the perspective of HOMFLY-PT polynomials.

\acknowledgments
The work of AM is supported in part by the Prime Minister’s Research Fellowship provided by the Ministry of Education, Government of India. The work of KC is supported in part by the South African Research Chairs Initiative (SARChI) of the Department of Science and Innovation and the National Research Foundation. PR would like to acknowledge SPARC/2019-2020/P2116/ project funding. PR would also like to acknowledge the ICTP’s Associate programme where some work on this project had happened during her visit as senior associate. 


\bibliographystyle{unsrt}
\bibliography{toruslinks.bib}


\end{document}